\documentclass[twocolumn,showpacs,preprintnumbers,amsmath,amssymb]{revtex4}

\usepackage{graphicx}
\usepackage{dcolumn}
\usepackage{bm}

\begin{document}
\preprint{APS/123-QED}
\title{On the Electrodynamics of Moving Particles in Gravitational Fields}
\author{Cl\'audio Nassif da Cruz\\
        (e-mail: {\bf cnassif@decea.ufop.br})}
 \altaffiliation{{\bf UFOP}: Universidade Federal de Ouro Preto- Instituto de Ci\^encias Exatas e Aplicadas (ICEA)-Rua 37/115, Loanda-
 Jo\~ao Monlevade-MG, Brazil.}

\date{\today}
\begin{abstract}
 The present work aims to search for an implementation of new symmetries in the space-time in order to enable us to find a
 connection between electrodynamics and gravitation, from where quantum principles naturally emerge. To do that, first of all we build a
 heuristic model of the electromagnetic nature of the electron so that the influence of the gravitational field on the electrodynamics of
 such moving particle leads us essentially to an elimination of the classical idea of rest by introducing the idea of a universal minimum
 limit of speed ($V$). Such a lowest limit $V$, being unattainable by the particles, represents a fundamental and preferred reference frame
 connected to a universal background field (a vacuum energy) that breaks Lorentz symmetry. So there emerges a new principle of symmetry in
 the space-time at the subatomic level for very low energies close to the background frame ($v\approx V$), providing a fundamental
 understanding for the uncertainty principle.
\end{abstract}

\maketitle
\section{\label{sec:level1} Introduction}

   In 1905, when Einstein\cite{1} criticized the existence of the luminiferous
 ether defended by Lorentz\cite{2}, Fitzgerald\cite{3} and Poincar\'e\cite{4},
 he\cite{1} solved the incompatibility between the laws of motion in the
 newtonian mechanics paradigm (Galileo's principle of addition of speed) and the
 laws of electric and magnetic fields of light (speed of light $c$) by using the
 following intuitive reasoning:

  ``If someone could move at the speed of light (c), the light ray would be
 standing still in relation to such an observer, based on Galilean principles of
 velocity addition. However, this made no sense for the electromagnetic theory
 (Maxwell equations) because, if it were possible for someone to stand still
 over the crest of a light ray wave, the electromagnetic wave would become
 stationary to such an observer. Naturally this would lead to the rupture of the
 space-time dynamic symmetries which comes from the oscillations of the
 electromagnetic fields of the wave''.

  Such an incompatibility was resolved by changing the newtonian theory by means of a
 correction which takes in consideration the speed of light ($c$) as the maximum and
 constant limit of speed in order to preserve the covariance of Maxwell's
 relativistic equations. So these equations mantain the same form for any
 reference frame.

  The speed of light in vacuum ($c$) is constant, namely it does not depend
 on the speed of the light source. Such a reasoning led Einstein to conclude
 that the idea of the luminiferous ether is not needed since the speed of light
 ($c$) is invariant. Therefore, due to the invariance of the speed of light,
  space and time are relative quantities, that is to say they vary in accordance with the
 reference frame; other than what was thought under the newtonian theory where the speed
 $c$ would change whereas space and time remained as absolute quantities.

  During the last 22 years of his life, Einstein attempted to bring the
 principles of Quantum Mechanics-QM (uncertainties) and Electromagnetism (EM)
 into the theory of gravitation (General Relativity-GR) by means of a unified
 field theory\cite{5}. Unfortunately his unification program was not successful
 in establishing a consistent theory between QM, EM and GR.

  Currently the string theories inspired by an old idea of Kaluza\cite{6} and Klein
\cite{7} regarding extra dimensions in the space-time have been prevailing in the
scenario of attempts to find a unified theory\cite{8,9,10}.

  Motivated by Einstein's ideas in a search for new fundamental symmetries in
 Nature, our main focus is to go back to that point of the old
 incompatibility between mechanics and electrodynamics by extending his reasoning
 in order to look for new symmetries that implement gravitation into electrodynamics
 of moving particles. We introduce more symmetries into the space-time where gravity and
electromagnetism become coupled to each other in order to build a new dynamics that provides
a fundamental understanding of the quantum uncertainties.

 Besides quantum gravity at the Planck length scale, our new symmetry idea appears due to the indispensable
presence of gravity at quantum level for particles with very large wavelengths (very low energies).
 This leads us to postulate a universal minimum speed related to a fundamental (privileged)
reference frame of background field that breaks Lorentz symmetry\cite{11}.

 Similarly to Einstein's reasoning, which has solved that old incompatibility
 between the nature of light and the motion of matter (massive objects), let us now expand it by making the following
heuristic assumption based on new symmetry arguments, namely:

  -In order to preserve the symmetry (covariance) of Maxwell's equations,
 the speed $c$ is required to be constant based on Einstein's reasoning, according to
 which it is forbidden to find the rest reference frame for the speed $c$
 due to the coexistence of the fields $\vec E$ and $\vec B$ in equal-footing. Now let us extend this reasoning
 by considering that $\vec E$ and $\vec B$ may also coexist for moving massive particles (as electrons), which are at
 subluminal level ($v<c$). So, by making such an assumption, it would be also
 impossible to find a rest reference frame for the massive particle by canceling
 its magnetic field, i.e., $\vec B=0$ with $\vec E\neq 0$. This would break the
 coexistence of these two fields, which would not be possible because we cannot find a
 reference frame at rest in such a space-time due to the minimum limit of speed $V$. Thus we should have
 $\vec E\neq 0$ and also $\vec B\neq 0$ for any change of reference frame due to the impossibility to find a null
 momentum for the electron, in a similar way for the photon.

 The hypothesis of the lowest non-null limit of speed for low energies ($v<<c$) in the space-time results in the
following physical reasonings:

- In non-relativistic quantum mechanics, the plane wave wave-function ($Ae^{\pm ipx/\hbar}$) which represents a free particle is an
 idealisation that is impossible to conceive under physical reality. In the event of such an idealized plane wave, it would be possible to
 find with certainty the reference frame that cancels its momentum ($p=0$) so that the uncertainty on its position would be $\Delta
 x=\infty$. However, the presence of an unattainable minimum limit of speed emerges in order to avoid the ideal case of a plane wave
 wave-function ($p=constant$ or $\Delta p=0$). This means that there is no perfect inertial motion ($v=constant$) such as a plane
 wave in QM, except the privileged reference frame of a universal background field connected to an unattainable minimum limit of speed $V$,
  where $p$ would vanish. However, since such a minimum speed $V$ (universal background frame) is unattainable for the particles with low
 energies (large length scales), their momentum cannot vanish when one tries to be closer to such a preferred frame ($V$). This is the
reason why we can never find a reference frame at rest where $\vec B=0$ for the charged particle in such a space-time.

 On the other hand, according to Special Relativity (SR), the momentum cannot be infinite since the maximum speed $c$ is also unattainable
 for a massive particle, except the photon ($v=c$) as it is a massless particle.

 This reasoning allows us to think that the electromagnetic radiation (photon:$``c-c"=c$) as well as the massive particle ($``v-v">V
(\neq 0)$ for $v<c$\cite{12}) are in equal-footing in the sense that it is not possible to find a reference frame at rest
 ($v_{relative}=0$) for both through any speed transformation in a space-time with a maximum and minimum limit of speed\cite{12}. Therefore
 such a deformed special relativity was denominated as Symmetrical Special Relativity (SSR)(see publication\cite{12}).

 The dynamics of particles in the presence of a universal (privileged) background reference frame connected to $V$ is
 within the context of ideas of Mach\cite{13}, Schr\"{o}dinger\cite{14} and Sciama\cite{15}, where there should be an absolute inertial
 reference frame in relation to which we have the inertia of all moving bodies. However, we must emphasize that the approach we intend to
 use is not classical as the machian ideas since the lowest limit of speed $V$ plays the role of a preferred reference frame
 of background field instead of the inertial frame of fixed stars.

 It is very interesting to notice that the idea of a universal background field was sought in vain by
 Einstein\cite{16}, motivated firstly by Lorentz. It was Einstein who coined the term {\it ultra-referential} as the fundamental aspect of
 reality for representing a universal background field\cite{17}. Based on such a concept, let us call {\it ultra-referential} $S_V$ to be
 the universal background field of a fundamental (preferred) reference frame connected to $V$.

 In the next section, a heuristic model will be built to describe the electromagnetic nature of the matter. It is based on Maxwell
 theory used for investigating the electromagnetic nature of a photon when the amplitudes of electromagnetic wave fields are normalized for
 one single photon with energy $\hbar w$. Thus, due to reciprocity and symmetry reasonings, we shall extend such a concept to the matter
 (e.g.: electron) through the idea of pair materialization after $\gamma$-photon decay.

\section{\label{sec:level1} Electromagnetic Nature of the Photon and of the Matter}

\subsection{Electromagnetic nature of the photon}

   In accordance with some laws of Quantum Electrodynamics\cite{18}, we may assume the electric field of a plane electromagnetic wave,
 whose amplitude is normalized for just one single photon\cite{18}. To do this, consider the vector potential of a plane
 electromagnetic wave, as follows:

\begin{equation}
\vec A=a cos(wt-\vec k.\vec r)\vec e,
\end{equation}
where $\vec k.\vec r=kz$, admitting that the wave propagates in the direction of z, $\vec e$ being the unitary vector of
polarization. Since we are in vacuum, we have

\begin{equation}
\vec E=-\frac{1}{c}\frac{\partial\vec A}{\partial{t}}=(\frac{wa}{c})sen(wt-kz)
\vec e
\end{equation}

  In the Gaussian System of units, we have $|\vec E|=|\vec B|$. So the average energy density of this wave shall be

\begin{equation}
\left<\rho_{eletromag}\right>=
\frac{1}{8\pi}\left<|\vec E|^2+|\vec B|^2\right>=
\frac{1}{4\pi}\left<|\vec E||\vec B|\right>,
\end{equation}
where $|\vec E||\vec B|=|\vec E|^2=|\vec B|^2$.

Inserting (2) into (3), we obtain

\begin{equation}
\left<\rho_{eletromag}\right>=
\frac{1}{8\pi}\frac{w^2a^2}{c^2},
\end{equation}
where $a$ is an amplitude which depends upon the number of photons.

  We wish to obtain the plane wave of one single photon. So, imposing this condition ($\hbar w$) in (4) and considering a
unitary volume for the photon ($v_{ph}=1$), we find

\begin{equation}
a=\sqrt{\frac{8\pi\hbar c^2}{w}}
\end{equation}

Inserting (5) into (2), we get

\begin{equation}
\vec{E}(z,t)=\frac{w}{c}\sqrt{\frac{8\pi\hbar c^2}{w}}sen(wt-kz)\vec e,
\end{equation}
from where we deduce

\begin{equation}
e_0=\frac{w}{c}\sqrt{\frac{8\pi\hbar c^2}{w}}=\sqrt{8\pi\hbar w},
\end{equation}
where $e_0$ could be thought of as an electric field amplitude normalized for one single photon, with $b_0=e_0$ (Gaussian system) being
the magnetic field amplitude normalized for just one photon. So we may write

\begin{equation}
\vec{E}(z,t)= e_0sen(wt-kz)\vec e
\end{equation}

Inserting (8) into (3) and considering the unitary volume ($v_{ph}=1$), we find

\begin{equation}
\left<E_{eletromag}\right>=\frac{1}{8\pi}e_0^2\equiv\hbar w
\end{equation}

Now, following the classical theory of Maxwell for the electromagnetic wave, let us consider an average quadratic electric field normalized
for one single photon, namely $e_m=e_0/\sqrt{2}=\sqrt{\left<|\vec E|^2\right>}$ (see (8)). So by doing this, we may write (9) in
the following alternative way:

\begin{equation}
\left<E_{eletromag}\right>=\frac{1}{4\pi}e_m^2\equiv\hbar w,
\end{equation}
where it happens

\begin{equation}
e_m=\frac{e_0}{\sqrt{2}}=\frac{w}{c}\sqrt{\frac{4\pi\hbar c^2}{w}}
=\sqrt{4\pi\hbar w}.
\end{equation}

   Here it is important to emphasize that, although the field given in (8) is normalized for only one photon, it is still a classical field
 of Maxwell in the sense that its value oscillates like a classical wave (solution (8)); the only difference here is that we have thought
 about a small amplitude field for just one photon.

   Actually the amplitude of the field ($e_0$) cannot be measured directly. Only in the
 classical approximation (macroscopic case), where we have a very large number of photons ($N\rightarrow\infty$), can we somehow measure
 the macroscopic field $\vec E$ of the wave. Therefore, although we could idealize the case of just one photon as if it were a Maxwell
 electromagnetic wave with small amplitude, the solution (8) is even a classical solution since the field $\vec E$ presents oscillation.

   On the other hand, we already know that the photon wave is a quantum wave, i.e., it is a de-Broglie wave where its wavelength
 ($\lambda=h/p$) is not interpreted classically as the oscillation frequency (wavelength due to oscillation) of a classical field because,
 if it were so, using the classical solution (8), we would have

\begin{equation}
E_{eletromag}=\frac{1}{4\pi}|\vec{E}(z,t)|^2=
\frac{1}{4\pi}e_0^2sen^2(wt-kz)
\end{equation}

    If the wave of a photon were really a classical wave, then its energy would not have a fixed value according to (12).
  Consequently, its energy $\hbar w$ would be only an average value [see (10)]. Hence, in order to achieve consistency between the result
 (10) and the quantum wave (de-Broglie wave), we must interpret (10) as being related to the de-Broglie wave of the
 photon with a discrete and fixed value of energy $\hbar w$ instead of an average energy value since we should consider the wave of
 one single photon being a non-classical wave, namely a de-Broglie wave. Thus we simply rewrite (10), as follows:

\begin{equation}
E_{eletromag}=E=pc=\frac{hc}{\lambda}=\hbar w\equiv\frac{1}{4\pi}e_{ph}^2,
\end{equation}
where we conclude

\begin{equation}
\lambda\equiv\frac{4\pi hc}{e_{ph}^2},
\end{equation}
 being $\lambda$ the de-Broglie wavelength. Now, according to (14), the single photon field $e_{ph}$ should not be assumed as a mean
 value for oscillating classical field, and we shall preserve it in order to interpret it as a {\it quantum electric field}, i.e.,
 a microscopic field of one photon. Let us also call it as a {\it scalar electric field} for representing the quantum mechanical
(corpuscular) aspect of the magnitude of electric field for one single photon. As the scalar field $e_{ph}$ is responsible for the energy
 of the photon ($E\propto e_{ph}^2$), where $w\propto e_{ph}^2$ and $\lambda\propto 1/e_{ph}^2$, we realize that $e_{ph}$ presents a
 quantum behavior since it provides the dual aspect (wave-particle) of the photon so that its mechanical momentum may be written as
 $p=\hbar k=2\pi\hbar/{\lambda}$=$\hbar e_{ph}^2/2hc$ [refer to (14)], or simply $p=e_{ph}^2/4\pi c$.

\subsection{The electromagnetic nature of the matter}

   Our goal is to extend the idea of photon electromagnetic energy [equation (13)] to the matter. By doing this, we shall provide
 heuristic arguments that rely directly on de-Broglie reciprocity postulate, which has extended the idea of wave (photon wave) to the
 matter (electron) also behaving like wave. Thus, the relation (14) for the photon, which is based on de-Broglie relation ($\lambda=h/p$)
 may be also extended to the matter (electron) in accordance with the very idea of de-Broglie reciprocity. In order to strengthen such
 argument, besides this, we are going to assume the phenomenon of pair formation where the $\gamma$-photon decays into two charged massive
 particles, namely the electron ($e^{-}$) and its anti-particle, the positron ($e^{+}$). Such an example will enable us to better
 understand the need of extending the idea of the photon electromagnetic mass ($m_{electromag} = E_{electromag}/c^2$: equation 13) to the
 matter ($e^{-}$ and $e^{+}$) by using the heuristic assumption about {\it scalar electromagnetic fields} for simply representing the
 magnitudes of such fields.

  Now consider the phenomenon of pair formation, i.e., $\gamma\rightarrow e^{-}+e^{+}$. Taking into account the conservation of energy for
$\gamma$-decay, we write the following equation:
\begin{equation}
E_{\gamma}=\hbar w = m_{\gamma}c^2=m_0^{-}c^2+m_0^{+}c^2 + K^{-}+ K^{+},
\end{equation}
 where $m_0^{-}c^2+m_0^{+}c^2=2m_0c^2$ since electron and positron have the same mass. $K^{-}$ and $K^{+}$ represent the kinetic energies
of the electron and the positron respectively. We have $m_0^{-}c^2=m_0^{+}c^2\cong 0,51MeV$.

Since the electromagnetic energy of the $\gamma$-photon is $E_{\gamma}=h\nu=m_{\gamma}c^2=\frac{1}{4\pi}e_{\gamma}^2=\frac{1}{4\pi}e_{\gamma}b_{\gamma}$,
or else, in IS (International System) of units we have $E_{\gamma}=\epsilon_0e_{\gamma}^2$, and also knowing that
 $e_{\gamma}=cb_{\gamma}$ in IS, where $b_{\gamma}$ is the {\it magnetic scalar field} of the $\gamma$-photon, we may also write

\begin{equation}
E_{\gamma}=c\epsilon_0(e_{\gamma})(b_{\gamma})
\end{equation}

  Photon has no charge, however when $\gamma$-photon is materialized into the pair electron-positron, its electromagnetic content given
 in (16) ceases to be free or purely kinetic (purely relativistic mass) to become massive due to the materialization of the pair. Since
 such massive particles ($v_{(+,-)}<c$) also behave like waves in accordance with the de-Broglie idea, it would be natural to extend the
 relation (14) of the photon for representing wavelengths of the matter (electron or positron) after $\gamma$-decay, namely:

\begin{equation}
\lambda_{(+,-)}\propto\frac{hc}{\epsilon_0[e_s^{(+,-)}]^2}=
\frac{h}{\epsilon_0[e_s^{(+,-)}][b_s^{(+,-)}]},
\end{equation}
where the fields $e_s^{(+,-)}$ and $b_s^{(+,-)}$ play the role of the electromagnetic contents of the energy, namely the
{\it scalar electromagnetic fields}. Thus, such scalar fields provide the total energies (masses) of the
 moving massive particles $e^{-}$ and $e^{+}$, being their masses essentially of electromagnetic origin given in the form, as follows:

\begin{equation}
m\equiv m_{electromag}\propto e_sb_s,
\end{equation}
where $E=mc^2\equiv m_{electromag}c^2$.

Using (16) and (17) as a basis, we may write (15) in the following way:

\begin{equation}
E_{\gamma}=c\epsilon_0e_{\gamma}b_{\gamma}=
c\epsilon_0e_s^{-}b_s^{-}v_e^{-}+c\epsilon_0e_s^{+}b_s^{+}v_e^{+},
\end{equation}
where $c\epsilon_0e_s^{-}b_s^{-}v_e^{-}=(c\epsilon_0e_{s0}^{-}b_{s0}^{-}v_e + K^{-})=(m_0^{-}c^2+ K^{-})$ and
$c\epsilon_0e_s^{+}b_s^{+}v_e^{+}=(c\epsilon_0e_{s0}^{+}b_{s0}^{+}v_e + K^{+})=(m_0^{+}c^2+ K^{+})$.

  The quantities $e_{s0}^{(+,-)}$ and $b_{s0}^{(+,-)}$ represent the proper scalar electromagnetic fields of the electron or positron.

   A fundamental point which the present heuristic model challenges is that, in accordance with equation (19), we realize that the electron
 is not necessarily an exact punctual particle. Quantum Electrodynamics, based on Special Relativity (SR), deals with the electron as a
 punctual particle. The well-known classical theory of the electron foresees for the radius of the electron the same order of magnitude of
 the proton radius, i.e., $R_e\sim 10^{-15}m$.

  The most recent experimental evidence about scattering of electrons by electrons at very high kinetic energies indicates that the
 electron can be considered approximately as a punctual particle. Actually the electrons have an extent less than collision distance, which
 is about $R_e\sim 10^{-16}m$\cite{19}. Of course such an extent is negligible in comparison to the dimensions of an atom ($10^{-10}m$) or
 even the dimensions of a nucleus ($10^{-14}m$), but it is not exactly punctual. By this reason, the present model can provide a very small
 non-null volume $v_e$ of the electron. But, if we just consider $v_e=0$ according to (19), we would have an absurd result, i.e,
  divergent scalar fields ($e_{s0}=b_{s0}\rightarrow\infty$). However, for instance, if we consider $R_e\sim 10^{-16}m$ ($v_e\propto
 R_e^3\sim 10^{-48}m^3$) in our model, and knowing that $m_0c^2\cong 0,51MeV (\sim 10^{-13}J)$, hence, in this case (see (19)), we would
 obtain $e_{s0}\sim 10^{23}V/m$.  This value is extremely high and therefore we may conclude that the electron is extraordinarily compact
  having a high mass (energy) density. If we imagine over the ``surface'' of the electron or even inside it, we would detect a
 constant and finite scalar field $e_{s0}\sim 10^{23}V/m$ instead of a divergent value for it. So according to the present model, the
 quantum scalar field $e_{s0}$ inside the almost punctual non-classical electron with radius $\sim 10^{-16}m$ would be finite and constant
 ($\sim10^{23}V/m$) instead of a function like $1/r^2$ with a divergent behavior. Of course, for $r>R_e$, we have the external vectorial
 (classical) field $\vec E$, decreasing with $1/r^2$, i.e, $E=e/r^2$ (see figure 1).

 The next section will be dedicated to the investigation about the electron coupled to a gravitational field according to the present
heuristic model.

\section{\label{sec:level1} A heuristic model for the electron coupled to gravity}

\subsection{Photon in a gravitational potential}

 When a photon with energy $h\nu$ is subjected to a gravitational potential $\phi$, its energy $E$ and frequency $\nu$ increase to
 $E^{\prime}=h\nu^{\prime}$, being

\begin{equation}
E^{\prime}=h\nu^{\prime}=h\nu\left(1+\frac{\phi}{c^2}\right)
\end{equation}

As, by convention, we have defined $\phi>0$ for an attractive potential, we get $\nu^{\prime}>\nu$. Considering the relation (16)
 given for any photon and inserting (16) into (20), we alternatively write

\begin{equation}
E^{\prime}=c\epsilon_0 e_{ph}^{\prime}b_{ph}^{\prime}=
c\epsilon_0 e_{ph}b_{ph}\sqrt{g_{00}},
\end{equation}
where $g_{00}$ is the first component of the metric tensor, being $\sqrt{g_{00}}=\left(1+\frac{\phi}{c^2}\right)$ and $e_{ph}=cb_{ph}$.

From (21), we can extract the following relations, namely:

\begin{equation}
e_{ph}^{\prime}=e_{ph}\sqrt{\sqrt{g_{00}}},~ ~
b_{ph}^{\prime}=b_{ph}\sqrt{\sqrt{g_{00}}}.
\end{equation}

  Due to the presence of gravity, the scalar fields $e_{ph}$ and $b_{ph}$ of the photon increase according to (22), leading to the
 increasing of the photon frequency or energy according to (20). Thus we may think about the increments of scalar fields in the
 presence of gravity, namely:

\begin{equation}
\Delta e_{ph}=e_{ph}(\sqrt{\sqrt{g_{00}}} - 1),~ ~
\Delta b_{ph}=b_{ph}(\sqrt{\sqrt{g_{00}}} - 1),
\end{equation}
being $\Delta e_{ph}=e_{ph}^{\prime}-e_{ph}$ and $\Delta b_{ph}=b_{ph}^{\prime}-b_{ph}$.

\subsection{Electron in a gravitational potential}

  When a massive particle with mass $m_0$ moves in a weak gravitational potential $\phi$, its total energy $E$ is

\begin{equation}
E=mc^2=m_0c^2\sqrt{g_{00}} + K,
\end{equation}
where we can think that $m_0(=m_0^{(+,-)})$ represents the mass of the electron (or positron) emerging from $\gamma$-decay in the presence
of a weak gravitational potential $\phi$.

In order to facilitate the understanding of what we are proposing, let us consider $K<<m_0c^2$ ($v<<c$) since we are interested only
in the influence of the potential $\phi$. Therefore, we simply write

\begin{equation}
E=m_0c^2\sqrt{g_{00}}.
\end{equation}

Since we already know that $E_0=m_0^{(+,-)}c^2=c\epsilon_0e_{s0}^{(+,-)}b_{s0}^{(+,-)}v_e$, we can also write the total energy $E$ as
follows:

\begin{equation} 
E =c\epsilon_0 e_s^{(+,-)}b_s^{(+,-)}v_e=
c\epsilon_0 e_{s0}^{(+,-)}b_{s0}^{(+,-)}v_e\sqrt{g_{00}},
\end{equation}
from where, we can extract

\begin{equation}
e_s^{(+,-)}=e_{s0}^{(+,-)}\sqrt{\sqrt{g_{00}}},~ ~
b_s^{(+,-)}=b_{s0}^{(+,-)}\sqrt{\sqrt{g_{00}}},
\end{equation}
in analogous way to (22).

So we obtain the following increments:

\begin{equation}
\Delta e_s=e_{s0}^{(+,-)}(\sqrt{\sqrt{g_{00}}} - 1), ~
\Delta b_s=b_{s0}^{(+,-)}(\sqrt{\sqrt{g_{00}}} - 1),
\end{equation}
where $\Delta e_s=c\Delta b_s$.

  As the energy of the particle can be represented as a kind of condensation of electromagnetic fields in the scalar forms with magnitudes
 $e_s$ and $b_s$, this heuristic model is capable of assisting us to think that the external fields $\vec{E}$ and $\vec{B}$
 of the moving charged particle, by storing an energy density ($\propto |\vec{E}|^2 +|\vec{B}|^2$), should also suffer perturbations
 (shifts) due to the presence of gravity (figure 1).

  We know that any kind of energy is also a source of gravitational field. This non-linearity that is inherent to a
 gravitational field leads us to think that the classical (external) fields $\vec E$ and $\vec B$ should suffer tiny shifts like
 $\delta\vec E$ and $\delta\vec B$ in the presence of a weak gravitational potential $\phi$. As such small shifts are positive having the
 same direction of $\vec E$ and $\vec B$, this should lead to a slight increasing of the electromagnetic energy density around the
 particle. And since the internal energy of the particle also increases in the presence of $\phi$ according to eq.(26), we expect that the
 magnitudes of the external shifts $\delta\vec E$ and $\delta\vec B$ should be proportional to the increments of internal (scalar) fields
 of the particle ($\Delta e_s$ and $\Delta b_s$), as follows:

\begin{equation}
\delta E\propto\Delta e_s=(e_s-e_{s0}),~ ~\delta B\propto\Delta b_s=(b_s-b_{s0}),
\end{equation}
being $\delta E=\delta E(\phi)=(E^{\prime}-E)>0$ and $\delta B=\delta B(\phi)=(B^{\prime}-B)>0$, where $\phi$ is the gravitational
potential. Here we have omitted the signs $(+,-)$ just for the purpose of simplifying the notation.

 In accordance with (29), we may conclude that there is a constant of proportionality that couples the external electromagnetic fields
$\vec E$ and $\vec B$ of the moving particle (electron) with gravity by means of the small shifts $\delta\vec E$ and $\delta\vec B$.
 So we write (29), as follows:

\begin{equation}
\delta\vec E=\vec\epsilon\xi\Delta e_s,~ ~
\delta\vec B=\vec\epsilon\xi\Delta b_s,
\end{equation}
where $\vec\epsilon$ is a unitary vector given in the same direction of $\vec E$ (or $\vec B$). So the small shift $\delta\vec E$
 (or $\delta\vec B$) has the same direction of $\vec E$ (or $\vec B$) (figure 1). The coupling $\xi$ is a dimensionaless proporcionality
 constant (a fine-tuning). We expect that $\xi<<1$ due to the fact that the gravitational interaction is much weaker than the
 electromagnetic one. The external shifts $\delta\vec E$ and $\delta\vec B$ depend only on gravitational potential $g_{00}$ over the
 electron (figure 1).

\begin{figure}
\begin{center}
\includegraphics[scale=0.65]{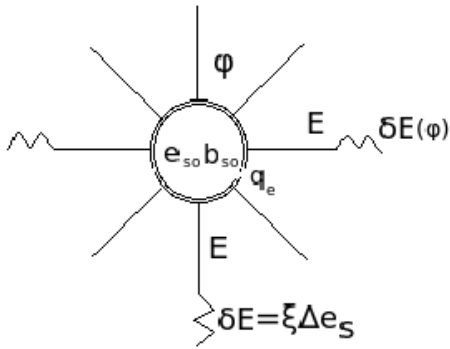}
\end{center}
\caption{This pictorial illustration shows us that the magnitude of the classical (external) field $\vec E$ of the particle suffers a
 small shift like $\delta\vec E(\phi)$ due to the gravitational potential $\phi$. The vector $\delta\vec E(\phi)$ has
the same direction of $\vec E$ so that the electric energy density around the charge increases in the presence of $\phi$. The magnitude
of the external tiny shift $\delta\vec E(\phi)$ (or $\delta\vec B(\phi)$) is proportional to the increment of internal (scalar) field $\Delta e_s$ (or $\Delta b_s$), i.e., $\delta E=\xi\Delta e_s$ (or $\delta B=\xi\Delta b_s$), being $\xi$ a tiny dimensionaless coupling
constant of gravi-electrical origin. The fields $e_{s0}$ and $b_{s0}$ are the internal (scalar) fields of the particle, i.e., they
represent the magnitudes of fields.}
\end{figure}

 Inserting (28) into (30), we obtain

\begin{equation}
\delta\vec E=\vec\epsilon\xi e_{s0}(\sqrt{\sqrt{g_{00}}}-1), ~ ~
\delta\vec B=\vec\epsilon\xi b_{s0}(\sqrt{\sqrt{g_{00}}}-1).
\end{equation}

 Due to the tiny positive shifts with magnitudes $\delta E$ and $\delta B$ in the presence of a gravitational potential $\phi$, the
 total electromagnetic energy density in the space around the charged particle is slightly increased, as follows:

\begin{equation}
\rho_{electromag}^{total}=\frac{1}{2}\epsilon_0[E+\delta E(\phi)]^2+\frac{1}{2\mu_0}
[B +\delta B(\phi)]^2.
\end{equation}

Inserting the magnitudes $\delta E$ and $\delta B$ from (31) into (32) and performing the calculations, we finally obtain

\begin{equation}
\rho_{total}=\frac{1}{2}[\epsilon_0 E^2+\frac{1}{\mu_0}B^2] +
\xi[\epsilon_0 Ee_{s0}+\frac{1}{\mu_0}Bb_{s0}](\sqrt{\sqrt{g_{00}}} - 1)+
\end{equation}~~~~~~~~~~~~~+$\frac{1}{2}\xi^2[\epsilon_0 (e_{s0})^2+
 \frac{1}{\mu_0}(b_{s0})^2](\sqrt{\sqrt{g_{00}}} - 1)^2$. \\

   We may assume that $\rho_{electromag}^{total}=\rho_{electromag}^{(0)}+\rho_{electromag}^{(1)}+\rho_{electromag}^{(2)}$
for representing (33), where $\rho_{electromag}^{(0)}$ is the free electromagnetic energy density (of zero order) for the ideal case
of a charged particle uncoupled from gravity ($\xi=0$), i.e, the ideal case of a free charge. We have $\rho^{(0)}\propto 1/r^4$
 ({\it coulombian term}).

  The coupling term $\rho^{(1)}$ (2nd.term) represents an electromagnetic energy density of first order since it contains a dependence
 of $\delta E$ and $\delta B$, i.e., it is proportional to $\delta E$ and $\delta B$ due to the influence of gravity.
 Therefore, it is a mixture term behaving essentially like a {\it radiation term}. So we find $\rho^{(1)}\propto 1/r^2$ as we have
 $E~(B)\propto 1/{r^2}$ and $e_{s0}~(b_{s0})\propto 1/r^0\sim constant$.  It is interesting to notice that this radiation term has origin
 from the non-inertial aspect of gravity that couples to the electromagnetic fields of the moving electric charge.

  The last coupling term ($\rho^{(2)}$) is purely interactive due to the presence of gravity only, namely it is a 2nd.order interactive
 electromagnetic energy density term since it is proportional to $(\delta E)^2$ and $(\delta B)^2$. And, as $e_{s0}~(b_{s0})\propto
 1/r^0\sim constant$, we find $\rho^{(2)}\propto 1/r^0\sim constant $, where we can also write $\rho^{(2)}=\frac{1}{2}\epsilon_0(\delta
 E)^2+\frac{1}{2\mu_0}(\delta B)^2$, which depends only on the gravitational potential ($\phi$) (see (31)).

  As we have $\rho^{(2)}\propto 1/r^0$, this term has a non-locality behavior. It means that
 $\rho^{(2)}$ behaves like a kind of non-local field that is inherent to the space (a term of background field). This term $\rho^{(2)}$ is
 purely from gravitational origin. It does not depend on the distance $r$ from the charged particle. Therefore $\rho^{(2)}$ is a uniform
 energy density for a given potential $\phi$ fixed on the particle.

   In reality, we generally have $\rho^{(0)}>>\rho^{(1)}>>\rho^{(2)}$. For a weak gravitational field, we can do a good practical
 approximation as $\rho_{eletromag}^{total}\approx\rho^{(0)}$; however, from a fundamental viewpoint, we cannot neglect the coupling terms
 $\rho^{(1)}$ and $\rho^{(2)}$, specially the last one $\rho^{(2)}$ for large distances since it has a vital importance in this work,
  allowing us to understand the constant energy density of background field, i.e., $\rho^{(2)}\propto 1/r^0\sim constant$. As $\rho^{(2)}$
 does not have $r$-dependence since $e_{s0}$ or $b_{s0}\sim constant$, it remains for $r\rightarrow\infty$. Section VII dedicates to
this question.

   The last term $\rho^{(2)}$ has deep implications for our understanding of the space-time structure at very large scales of length.
  In a previous paper\cite{12}, we had the opportunity to investigate the implications of the energy density $\rho^{(2)}$ in
 cosmology (a vacuum energy connected to a preferred background frame of an invariant minimum speed), where we have achieved
 interesting results regarding the problem of the accelerated expansion of the Universe.

   In the next section, we will estimate the coupling constant $\xi$ and consequently the idea of a universal minimum speed $V$ in the
 subatomic world.

   In a coming article we will investigate how the fields $\vec E$ and $\vec B$ ($F^{\mu\nu}$) transform with changing of reference frames
 for such space-time with a minimum speed. And, in the last section of this paper, we will investigate how the shifts
 $\delta\vec E$ and $\delta\vec B$ transform with the speed $v$ in SSR.

   In a previous publication\cite{12}, we have already verified that the invariant minimum speed $V$ connected to the preferred frame $S_V$
 of a background field does not affect the covariance of the Maxwell wave equations. Others of its implications should be also
 investigated elsewhere.

\section{\label{sec:level1} The fine adjustment constant $\xi$ and the minimum speed $V$}

   Let us begin this section by considering the well-known problem that deals with the electron in the bound state of a coulombian
 potential of a proton (Hydrogen atom). We have started from this issue because it poses an important analogy with the present
 model of the electron coupled to a gravitational field.

   We know that the {\it fine structure constant} ($\alpha_F\cong 1/137$) plays an important role for obtaining the energy levels that
bind the electron to the nucleus (proton) in the Hydrogen atom. Therefore, in a similar way to such idea, we plan to extend it in order to realize that the fine coupling constant $\xi$ plays an even more fundamental role than the fine structure $\alpha_F$ since the constant
 $\xi$ couples gravity to the electromagnetic field of the moving charge $q_e$ (PS: the spin of electron is not considered in this model).

  Let's initially consider the energy that binds the electron to the proton at the fundamental state of the Hydrogen atom, as follows:

\begin{equation}
\Delta E=\frac{1}{2}\alpha_F^2m_0c^2,
\end{equation}
where $\Delta E$ is assumed as module. We have $\Delta E<<m_0c^2$, where $m_0$ is the electron mass, which is close to the reduced mass
 of the system ($\mu\approx m_0$) since the mass of the proton is $m_p>>m_0$, being $m_p\approx 1840m_0$.

 We have $\alpha_F=e^2/\hbar c=q_e^2/4\pi\epsilon_0\hbar c\approx1/137$ (fine structure constant). Since $m_0c^2\cong 0,51 MeV$, from (34)
 we get $\Delta E\approx 13,6 eV$.

 As we already know that $E_0=m_0c^2=c\epsilon_0e_{s0}b_{s0}v_e$, we may write (34) in the following alternative way:

\begin{equation}
\Delta E=\frac{1}{2}\alpha_F^2c\epsilon_0e_{s0}b_{s0}v_e\equiv\frac{1}{2}c\epsilon_0(\Delta e_{s})(\Delta b_{s})v_e,
\end{equation}
from where we extract

\begin{equation}
\Delta e_{s}\equiv\alpha_F e_{s0};~~
\Delta b_{s}\equiv\alpha_F b_{s0}.
\end{equation}

 It is interesting to observe that the relations (36) maintain a certain analogy with (30); however, first of all we must emphasize that
 the variations (increments) $\Delta e_s$ and $\Delta b_s$ on the electron energy (given in 36) have purely coulombian origin since the
 fine structure constant $\alpha_F$ depends solely on the electron charge. Thus we could express the electric force between these two
 electronic charges in the following way:

\begin{equation}
F_e=\frac{e^2}{r^2}=\frac{q_e^2}{4\pi\epsilon_0 r^2}=\frac{\alpha_F
\hbar c}{r^2}.
 \end{equation}

 If we now ponder about a gravitational interaction between these two electrons, thus, in a similar way to (37), we have

\begin{equation}
F_g=\frac{Gm_e^2}{r^2}=\frac{\beta_F\hbar c}{r^2},
 \end{equation}
where we extract

\begin{equation}
\beta_F=\frac{Gm_e^2}{\hbar c}.
 \end{equation}

  We have $\beta_F<<\alpha_F$ due to the fact that the gravitational interaction is very weak when compared with the electrical
 interaction, so that $F_g/F_e=\beta_F/\alpha_F\sim 10^{-42}$, where $\beta_F\cong 1,75\times 10^{-45}$. Therefore we shall denominate
 $\beta_F$ as the {\it superfine structure constant} since the gravitational interaction creates a bonding energy extremely smaller than
 the coulombian one given for the fundamental state ($\Delta E$) of the Hydrogen atom.

 To sum up, we say that, whereas $\alpha_F(e^2)$ provides the adjustment for the coulombian bonding energies between two electronic
 charges, $\beta_F(m_e^2)$ gives the adjustment for the gravitational bonding energies between two electronic masses. Such bonding energies
 of electrical or gravitational origin lead to an increment of the energy of the particle by means of variations $\Delta e_s$
 and $\Delta b_s$.

   Following the above reasoning, we realize that the present model enables us to consider $\xi$ as the fine-tuning (coupling) between
 a gravitational potential created for instance by the electron mass $m_e$ and the electrical field (electrical energy density) created
 by another charge $q_e$ of its neighbor. Hence, in this more fundamental case, we have a kind of bond of the type
 ``$m_eq_e$" (mass-charge) given by the tiny coupling $\xi$.

   The way we follow for obtaining $\xi$ starts from important analogies by considering the ideas of fine structure constant
 $\alpha_F=\alpha_F(e^2)$ (electric interaction) and superfine structure constant $\beta_F=\beta_F(m_e^2)$ (gravitational interaction), so
 that it is easy to conclude that the kind of mixing coupling we are proposing here, of the type ``$m_eq_e$'', represents the
 gravi-electrical coupling constant $\xi$, namely $\xi$ is of the form $\xi=\xi(m_eq_e)$. So we get

\begin{equation}
\xi=\sqrt{\alpha_F\beta_F}
\end{equation}

As we already have $\alpha_F$ and $\beta_F$ (given in (39)), from (40) we finally obtain

\begin{equation}
\xi=\sqrt{\frac{G}{4\pi\epsilon_0}}\frac{m_eq_e}{\hbar c}=\frac{\sqrt{G}m_ee}
{\hbar c},
\end{equation}
where indeed we verify $\xi\propto m_eq_e$. So, from (41) we find $\xi\cong 3,57\times 10^{-24}$. Let us denominate $\xi$ as the {\it fine
 adjustment constant}.  We have $e=q_e/\sqrt{4\pi\epsilon_0}$. The quantity $\sqrt{G}m_e$ can be thought as if it were a {\it gravitational
 charge}.

  In the Hydrogen atom, we have the fine structure constant $\alpha_F=e^2/\hbar c=v_B/c$, where $v_B=e^2/\hbar\cong c/137$. This is the
 speed of the electron at the fundamental atomic level (Bohr velocity). At this level, the electron does not radiate because it is in a
 kind of balance state in spite of its electrostatic interaction with the nucleus (centripete force), that is to say it works as if it
 were effectively an inertial system.  So now by considering an analogous reasoning applied to the case of the gravi-electrical coupling
 constant $\xi$, we may write (41), as follows:

\begin{equation}
\xi=\frac{V}{c},
\end{equation}
from where we get

\begin{equation}
V=\sqrt{\frac{G}{4\pi\epsilon_0}}\frac{m_eq_e}{\hbar},
\end{equation}
being $V\cong 1,071\times 10^{-15}m/s$. This is a fundamental constant of Nature as well as the speed
 $c=q_e^2/4\pi\epsilon_0\alpha_{F}\hbar$.

  Similarly to the Bohr velocity $v_B(=\alpha_{F}c)$, the speed $V(=\xi c)$ is also a universal fundamental constant; however the
 crucial difference between them is that $V$ is associated with the most fundamental bound state in the Universe (a background energy)
 since gravity ($G$), which is the weakest interaction, plays now a fundamental role for the dynamics of the electron
 (electrodynamics) at low energies by means of the tiny gravi-electrical coupling $\xi=V/c\cong 3.57\times 10^{-24}$ (see eq.(33)).

  We aim to postulate $V$ as an unattainable universal minimum speed for all the particles in the subatomic world, but before this, we must
 provide a better justification of why we consider the electron mass and charge to calculate $V$ $(V\propto m_eq_e)$, instead of
 the masses and charges of other particles. Although there are fractionary electric charges such as is the case of quarks, such charges are
 not free in Nature for coupling only with gravity. They are strongly bonded by the strong force (gluons). Therefore, the charge of
 the electron is the smallest free charge in Nature. On the other hand, the electron is the elementary charged particle with the smallest
 mass. Consequently, the product $m_eq_e$ assumes a minimum value. In addition, the electron is completely stable. Other charged particles
 such as for instance $\pi^{+}$ and $\pi^{-}$ have masses that are much greater than the electron mass and therefore they are unstable
 decaying very quickly.

 Now we can verify that the minimum speed ($V$) given in (43) is directly related to the minimum length of quantum gravity
 (Planck length), as follows:

\begin{equation}
V=\frac{\sqrt{G}m_e e}{\hbar}=(m_ee\sqrt{\frac{c^3}{\hbar^3}})l_p, 
\end{equation}
where $l_p=\sqrt{G\hbar/c^3}$.

 In (44), since $l_p$ is directly related to $V$, if we make $l_p\rightarrow 0$, this implies $V\rightarrow 0$ and we recover the
 classical case of the space-time in SR. So we also recover the classical result from (33), i.e, $\rho^{total}_{electromag}=\rho^{(0)}$
 for $\xi=0$ (ideal case of the free electron uncoupled from gravity).

  The universal constant of minimum speed $V$ given in (44) for very low energy scales (very large wavelengths) is
 directly related to the universal constant of minimum length $l_p$ (very high energies), whose invariance has been studied
 in DSR (Double Special Relativity) by Magueijo, Smolin, Camelia et al\cite{29,30,31,32,33,34}.

 This research which redeems some features of those non-conventional ideas of Einstein regarding the introduction of a new concept of
 ``ether'' (a relativistic ``ether")\cite{21,22,23,24,25,26}, namely a background field (a vacuum energy) for the physical space, seeks to
 implement naturally the quantum principles\cite{25,26,27} in the space-time of SSR to be dealt with in the next two sections.

\section{\label{sec:level1} Reference frames and space-time interval in SSR}

   Before we deal with the implications due to the implementation of the ultra-referential $S_V$ in the space-time of SSR, let us make a
 brief presentation of the meaning of the Galilean reference frame (reference space), well-known in SR. In accordance with SR, when an
 observer assumes an infinite number of points at rest in relation to himself, he introduces his own reference space $S$. Thus,
 for another observer $S^{\prime}$ who is moving at a speed $v$ in relation to $S$, there should also exist an infinite number of points at
 rest in his own reference space. Therefore, for the observer $S^{\prime}$, the reference space $S$ is not standing still and has its
 points moving at a speed $-v$. It is for this reason that there is no privileged Galilean reference frame at absolute rest according to
 the principle of relativity, since the rest reference space of a given observer becomes motion to another one.

   The absolute space of pre-Einsteinian physics, connected to the ether in the old sense, also constituted by itself a reference space.
  Such a space was assumed as the privileged reference space of the absolute rest. However, as it was also essentially a Galilean reference
 space like any other, comprised of a set of points at rest, actually it was also subject to the notion of movement. The idea of movement
 could be applied to the ``absolute space'' when, for example, we assume an observer on Earth, which is moving at a speed $v$ in relation
 to such space. In this case, for an observer at rest on Earth, the points that would constitute the absolute space of reference would be
 moving at a speed of $-v$. Since the absolute space was connected to the old ether, the Earth-bound observer should detect a flow of ether
$-v$; however the Michelson-Morley experience has not substantiated the luminiferous ether that was denied by Einstein.

      In 1916, after the final formulation of GR, Einstein proposed new concepts of ``ether"\cite{21}\cite{22}\cite{28}. The new ``ether"
 was a kind of relativistic ``ether" (a background field) that described space-time as a {\it sui generis} material medium, which in no way
 could constitute a reference space subject to the relative notion of movement.

     Since we cannot think about a reference system constituted by a set of infinite points at rest in the quantum space-time of
 SSR\cite{12}, we should define a non-Galilean reference system essentially as a set of all the particles having the same state of movement
 (speed $v$) with respect to the ultra-referential $S_V$ (background frame) so that $v>V$. Hence, SSR should contain 3 postulates, namely:

 1)-the constancy of the speed of light ($c$);

 2)-the non-equivalence (asymmetry) of the non-Galilean reference frames due to the background frame $S_V$ that breaks Lorentz symmetry,
  i.e., we cannot exchange $v$ for $-v$ by the inverse transformations, since $``v-v">V$\cite{12};

 3)-the covariance of a relativistic ``ether" (a vacuum energy of the ultra-referential $S_V$) connected to an invariant minimum limit of
 speed $V$.

Let us assume the reference frame $S^{\prime}$ with a speed $v$ in relation to the ultra-referential $S_V$ according to figure 2.

\begin{figure}
\includegraphics[scale=0.65]{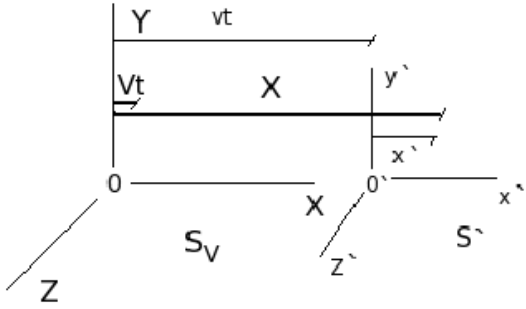}
\caption{$S^{\prime}$ moves with a speed $v$ with respect to the background field of the covariant ultra-referential $S_V$.
 If $V\rightarrow 0$, $S_V$ is eliminated (no vacuum energy) and thus the Galilean frame $S$ takes place, recovering Lorentz
 transformations.}
\end{figure}

Hence, consider the motion at one spatial dimension, namely $(1+1)D$ space-time with background field-$S_V$. So we write the following
 transformations:

  \begin{equation}
 x^{\prime}=\Psi(X-\beta_{*}ct)=\Psi(X-vt+Vt),
  \end{equation}
where $\beta_{*}=\beta\epsilon=\beta(1-\alpha)$, being $\beta=v/c$ and $\alpha=V/v$, so that $\beta_{*}\rightarrow 0$ for $v\rightarrow V$
 or $\alpha\rightarrow 1$.

 \begin{equation}
 t^{\prime}=\Psi(t-\frac{\beta_{*}X}{c})=\Psi(t-\frac{vX}{c^2}+\frac{VX}{c^2}),
  \end{equation}
being $\vec v=v_x{\bf x}$. We have $\Psi=\frac{\sqrt{1-\alpha^2}}{\sqrt{1-\beta^2}}$. If we make $V\rightarrow 0$ ($\alpha\rightarrow 0$),
 we recover Lorentz transformations where the ultra-referential $S_V$ is eliminated and simply replaced by the Galilean frame $S$ at rest
 for the classical observer. The above transformations and some of their implications were treated in a previous paper (see reference
 \cite{12}). In a further work, we should investigate whether such transformations form a group. Transformations in $(3+1)D$ will be also
 investigated elsewhere.

    According to SR, there is no ultra-referential $S_V$, i.e., $V\rightarrow 0$. So the starting point for observing $S^{\prime}$ is the
 reference space $S$ at which the classical observer thinks to be at rest (Galilean reference frame $S$).

   According to SSR, the starting point for obtaining the true motion of all the particles at $S^{\prime}$ is the ultra-referential
 $S_V$ (see figure 2). However, due to the non-locality of $S_V$, being unattainable by any particle, the existence of a classical
 observer at $S_V$ becomes inconceivable. Hence, let us think about a non-Galilean frame $S_0$ with a certain intermediary speed
 ($V<<v_0<<c$) in order to represent the new starting point (at local level) for observing the motion of $S^{\prime}$. At this
 non-Galilean reference frame $S_0$ (for $v=v_0$), which plays the similar role of a ``rest'', we must restore all the newtonian parameters
 of the particles such as the proper time interval $\Delta\tau$, i.e., $\Delta\tau=\Delta t$ for $v=v_0$; the mass $m_0$, i.e.,
 $m(v=v_0)= m_0$\cite{12}, among others. Therefore, in this sense, the frame $S_0$ under SSR plays a role that is similar to the frame $S$
 under SR where $\Delta\tau=\Delta t$ for $v=0$, $m(v=0)=m_0$, etc. However we must stress that the classical relative rest ($v=0$) should
 be replaced by a universal ``quantum rest'' $v_0(\neq 0)$ of the non-Galilean reference frame $S_0$ in SSR. We will show that $v_0$
 is also a universal constant. 

   Considering the improper non-Galilean reference frame $S_0$ so that $V~(S_V)<<v_0~(S_0)<<c$, we get figure 3.

\begin{figure}
\includegraphics[scale=0.55]{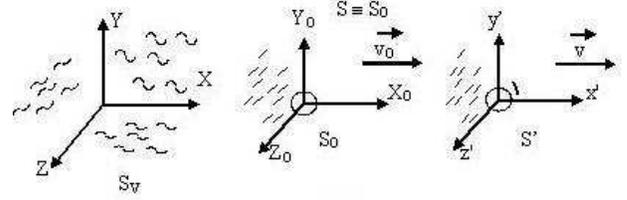}
\caption{As $S_0$ is fixed (universal), being $v_0 (>>V)$ given with respect to $S_V$, we should also consider the interval $V~(S_V)<v\leq
 v_0$. Such interval introduces a new symmetry for the space-time of SSR. Thus we expect that new and interesting results take place. For
 this interval, we have $\Psi(v)\leq 1$.}
\end{figure}

  In general we should have the interval of speeds $V<v<c$ assumed by $S^{\prime}$ (figures 2 and 3). According to figure 3, we notice that
 both $S_V$ and $S_0$ are already fixed or universal non-Galilean frames, whereas $S^{\prime}$ is a rolling non-Galilean frame of the
 particle moving within the interval of speeds $V<v<c$.

  Since the rolling frame $S^{\prime}$ is a non-Galilean frame due to the impossibility to find a set of points at rest on it, we cannot
 place a particle exactly on its origin $O^{\prime}$ as there should be a delocalization (an intrinsic uncertainty) $\Delta
 x^{\prime}$ ($=\overline{O^{\prime}C}$) around the origin $O^{\prime}$ of $S^{\prime}$ (see figure 4). Actually we want to show that such
 delocalization $\Delta x^{\prime}$ is a function which should depend on the speed $v$ of $S^{\prime}$ with respect to $S_V$, namely, for
 example, if $S^{\prime}\rightarrow S_V$ ($v\rightarrow V$), we would have $\Delta x^{\prime}\rightarrow\infty$ (complete delocalization)
 due to the non-local aspect of the ultra-referential $S_V$. On the other hand, if $S^{\prime}\rightarrow S_c(v\rightarrow c)$, we would
 have $\Delta x^{\prime}\rightarrow 0$ (much better located on $O^{\prime}$). So let us search for the function $\Delta x^{\prime}=\Delta
 x^{\prime}_v=\Delta x^{\prime}(v)$, starting from figure 4.

\begin{figure}
\includegraphics[scale=0.6]{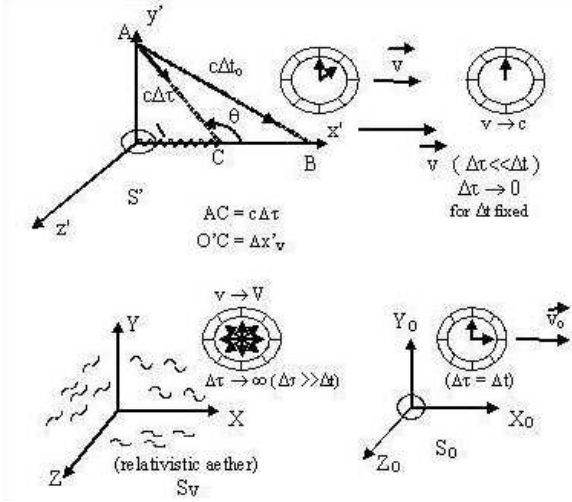}
\caption{We have four imaginary clocks associated to the non-Galilean reference frames $S_0$, $S^{\prime}$, the ultra-referential $S_V$ for
 $V$ and $S_c$ for $c$. We observe that the time (interval $\Delta\tau$) elapses much faster closer to infinite
 ($\Delta\tau\rightarrow\infty$) when one approximates to $S_V$ and, on the other hand, it tends to stop ($\Delta\tau\rightarrow 0$) when
 $v\rightarrow c $, providing a strong symmetry for SSR.}
\end{figure}

 At the reference frame $S^{\prime}$ in figure 4, let us consider a photon emitted from a point $A$ at $y^{\prime}$, in the direction
 $\overline {AO^{\prime}}$, which occurs only if $S^{\prime}$ were Galilean (at rest on itself). However, since the electron cannot be
 thought of as a point at rest on its proper non-Galilean reference frame $S^{\prime}$ and cannot be located exactly on $O^{\prime}$, then
 its delocalization $\overline {O^{\prime}C}$ ($=\Delta x^{\prime}_v$) causes the photon to deviate from the direction $\overline
 {AO^{\prime}}$ to $\overline {AC}$.  Hence, instead of simply the segment $\overline {AO^{\prime}}$, a rectangle triangle $AO^{\prime}C$
 is formed at the proper non-Galilean frame $S^{\prime}$ where it is not possible to find a set of points at rest.

   At the frame $S_0$ (``quantum rest'' $v_0$), which plays a similar role of the improper Galilean frame $S$ ($v=0)$, and from where one
 ``observes'' the speed $v$ of $S^{\prime}$ with respect to the background frame $S_V$, we see the path $\overline {AB}$ of the photon
 (figure 4). Hence the rectangle triangle $AO^{\prime}B$ is formed. Since the vertical leg $\overline {AO^{\prime}}$ is common to the
 triangles $AO^{\prime}C$ ($S^{\prime}$) and $AO^{\prime}B$ ($S_0$), we have

 \begin{equation}
(\overline {AO^{\prime}})^2=(\overline {AC})^2-(\overline {O^{\prime}C})^2=
(\overline {AB})^2 - (\overline {O^{\prime}B})^2,
\end{equation}
 that is

 \begin{equation}
(c\Delta\tau)^2-(\Delta x^{\prime}_v)^2=(c\Delta t)^2 - (v\Delta t)^2,
\end{equation}
where $\Delta t=\Delta t_0$ (improper time at $S_0$), being $S_0$ the improper non-Galilean frame in SSR. So, from (48) we expect that, for the case $v=v_0$ (``quantum rest" $S_0$), we have $\Delta\tau=\Delta t$, leading to $\Delta x^{\prime}_{[v=v_0]}=v_0\Delta\tau$ (see (48)).

  Now we will search for $\Delta x^{\prime}_v$. If $\Delta x^{\prime}(v)=\Delta x^{\prime}_v=0$, we fall back to the classical case (SR)
 where we consider, for instance, a train wagon ($S^{\prime}$) moving in relation to a fixed rail ($S$). At a point A on the ceiling of the
 wagon, there is a laser which releases photons toward $y^{\prime}$, reaching the point $O^{\prime}$ assumed on the origin (on the
 floor). For the proper Galilean frame $S^{\prime}$, the trajectory of the photon is simply the segment $\overline {AO^{\prime}}$. For the
 improper Galilean frame $S$ at rest, its trajectory is $\overline {AB}$.

  As $\Delta x^{\prime}_v$ is a function of $v$, being an ``internal displacement" (delocalization) given on the proper non-Galilean
 frame $S^{\prime}$, we may write it in the following way:

\begin{equation}
\Delta x^{\prime}_v= f(v)\Delta\tau,
\end{equation}
 where $f(v)$ is a function of $v$. It has also dimension of speed, however it could be thought as if it were a kind of
{\it ``internal motion" $v_{int}$} given in a reciprocal space of momentum for representing the delocalization $\Delta x^{\prime}_v$
on position of the particle at its proper non-Galilean reference frame $S^{\prime}$, i.e., $\Delta x^{\prime}_v=v_{int}\Delta\tau$,
 where $f(v)=v_{int}$.

  In figure 4 we can see that such a proper delocalization $\Delta x^{\prime}_v$ is given by the segment $\overline{O^{\prime}C}$ at
the frame $S^{\prime}$, i.e., $\Delta x^{\prime}_v=\overline{O^{\prime}C}$. This leads us to think that there is effectively an intrinsic
uncertainty $\Delta x^{\prime}_v$ on position of the particle, as we will see later (section VI). So inserting (49) into (48), we obtain

\begin{equation}
\Delta\tau\left[1-\frac{[f(v)]^2}{c^2}\right]^{\frac{1}{2}}=\Delta t\left(1-\frac{v^2}{c^2}\right)^{\frac{1}{2}},
\end{equation}
 where $f(v)=v_{int}=v_{reciprocal}=v_{rec}$.

  As we have $v\leq c$, we should find $f(v)\leq c$ in order to avoid an imaginary number in the 1st. member of (50).

 The domain of $f(v)$ is such that $V\leq v\leq c$. Thus let us also think that its image is $V\leq f(v)\leq c$ since $f(v)$ has dimension
 of speed for representing $v_{int}$, which also should be limited by $V$ and $c$.

  Let us make $[f(v)]^2/c^2 = f^2/c^2 =v_{int}^2/c^2=\alpha^2$, whereas we already know that $v^2/c^2 =\beta^2$. Thus, from (50) we
have the following cases:

- (i) When $v\rightarrow c$ ($\beta\rightarrow\beta_{max}=1$), the relativistic correction in its 2nd member (right-hand side) prevails,
 while the correction on the left-hand side becomes practically neglectable, i.e., we should have $v_{int}=f(v)<<c$, where
 $lim._{v\rightarrow c}f(v)=f_{min}=(v_{int})_{min}=V$ ($\alpha\rightarrow\alpha_{min}=V/c=\xi$).

 -(ii) On the other hand, due to the idea of symmetry, if $v\rightarrow V$ ($\beta\rightarrow\beta_{min}=V /c =\xi$), there is no
 substantial relativistic correction on the right-hand side of (50), while the correction on the left-hand side becomes now considerable,
 that is, we should have $lim._{v\rightarrow V}f(v)=f_{max}=(v_{int})_{max}=c$ ($\alpha\rightarrow\alpha_{max}=1$).

 In short, from (i) and (ii) we observe that, if $v\rightarrow v_{max}=c$, then $f\rightarrow f_{min}=(v_{int})_{min}=V$ and, if
 $v\rightarrow v_{min}=V$, then $f\rightarrow f_{max}=(v_{int})_{max}=c$. So now we indeed perceive that the ``internal motion" $v_{int}$
 works like a reciprocal speed $v_{rec}$ leading to the delocalization $\Delta x^{\prime}_v$ on position. In other words, we say that, when
 the speed $v$ increases to $c$, the reciprocal one ($v_{rec}$) decreases to $V$. On the other hand, when $v$ tends to $V$ ($S_V$), so
 $v_{rec}$  tends to $c$ leading to a very large delocalization $\Delta x^{\prime}_v$. Due to this fact, we reason that

\begin{equation}
f(v)=v_{int}=v_{rec}=\frac{a}{v},
\end{equation}
where $a$ is a constant that has dimension of squared speed. Such reciprocal speed $v_{rec}$ will be better understood
later. It is interesting to notice that an almost similar idea of considering an ``internal motion" for microparticles was also thought by
Natarajan\cite{35} in order to try to introduce a connection between SR and MQ.

In addition to (50) and (51), we already know that, at the non-Galilean frame $S_0$, we should have the condition of equality of the time
 intervals, namely $\Delta t=\Delta\tau$ for $v=v_0$. In accordance with (50), this occurs only if

\begin{equation}
\frac{[f(v_0)]^2}{c^2}=\frac{v_0^2}{c^2}\Rightarrow f(v_0)=v_0.
\end{equation}

 Inserting the condition (52) into (51), we find

\begin{equation}
a=v_0^2
\end{equation}

  And so we obtain

\begin{equation}
f(v)=v_{int}= v_{rec}=\frac{v_o^2}{v}
\end{equation}

According to (54) and also considering the cases (i) and (ii), we observe respectively that $f(c)=V=v_0^2/c$ ($V$ is the reciprocal
speed of $c$) and $f(V)=c=v_0^2/V$ ($c$ is the reciprocal speed of $V$), from where we find

\begin{equation}
v_0=\sqrt{cV}
\end{equation}

As we already know the value of $V$ (refer to (43)) and $c$, we compute the speed of ``quantum rest'' $v_0\cong 5.65\times
 10^{-4}m/s$, which is universal because it depends on the universal constants $c$ and $V$. However we must stress that only $c$ and $V$
 remain invariant under velocity transformations in SSR\cite{12}.

Finally, by inserting (55) into (54) and after into (50), we finally obtain

\begin{equation} 
\Delta\tau\sqrt{1-\frac{\it{V}^2}{v^2}}=\Delta t\sqrt{1-\frac{v^2}{c^2}},
\end{equation}
being $\alpha=f(v)/c=v_{rec}/c=V/v$ and $\beta=v/c$. In fact, if $v=v_0=\sqrt{cV}$ in (56), then we have $\Delta\tau =\Delta t$. Therefore,
 we conclude that $S_0(v_0)$ is the improper non-Galilean reference frame of SSR, so that, if

a) $v>>v_0$ ($v\rightarrow c$)$\Rightarrow\Delta t>>\Delta\tau$: It is the well-known {\it improper time dilatation} of SR.

b) $v<<v_0$ ($v\rightarrow V$) $\Rightarrow\Delta t<<\Delta\tau$: Let us call this new result {\it contraction of improper time}. This
 shows us the novelty that the proper time interval ($\Delta\tau$) may dilate in relation to the improper one ($\Delta t$), being
 $\Delta\tau$ a variable of time that is intrinsic to particle on its proper non-Galilean reference frame $S^{\prime}$. Such effect would
 become more evident only if $v\rightarrow V$, as we would have $\Delta\tau\rightarrow\infty$. In other words, this means that the
 proper time $\tau$ ($S^{\prime}$) would elapse much faster than the improper one $t$ ($S_0$).

  In SSR it is interesting to notice that we recover the newtonian regime of speeds only if $V<<v<<c$, which represents an intermediary
 regime where $\Delta\tau\approx\Delta t$.

  Substituting (54) in (49) and also considering (55), we obtain

\begin{equation}
\overline{O^{\prime}C}=\Delta x^{\prime}_v=v_{rec}\Delta\tau=\frac{v_0^2}{v}\Delta\tau=\frac{cV}{v}\Delta\tau=\alpha c\Delta\tau
\end{equation}

  We verify that, if $V\rightarrow 0$ or $v_0\rightarrow 0\Rightarrow\overline {O^{\prime}C} = \Delta x^{\prime}_v=0$,
 restoring the classical case of SR where there is no motion in reciprocal space, i.e., $v_{rec}=0$. And also, if $v>>v_0\Rightarrow\Delta
 x^{\prime}_v\approx 0$, i.e., we get an approximation where $v_{rec}$ can be neglected. We also verify that, if $v=v_0$, this implies
$\Delta x^{\prime}_{[v=v_0]}=v_0\Delta\tau$, as we have already obtained from (48) with the condition $\Delta t=\Delta\tau$ ($v=v_0$).
 This is the unique condition where we find $v=v_{rec}=v_0$.

  From (57), it is also important to notice that, if $v\rightarrow c$, we have $\Delta x^{\prime}(c)=V\Delta\tau$ and, if $v\rightarrow V$
 ($S_V$), we have $\Delta x^{\prime}(V)=c\Delta\tau$. This means that, when the particle momentum increases ($v\rightarrow c$), such a
 particle becomes much better localized upon itself ($V\Delta\tau\rightarrow 0$); and when its momentum decreases ($v\rightarrow V$), it
 becomes completely delocalized because it gets closer to the non-local ultra-referential $S_V$ where $\Delta
 x^{\prime}_v=\Delta x^{\prime}_{max}=c\Delta\tau\rightarrow\infty$. That is the reason why we realize that speed $v$ (momentum) and
 position (delocalization $\Delta x^{\prime}_v=v_{rec}\Delta\tau$) operate like mutually reciprocal quantities in SSR since we have
 $\Delta x^{\prime}_v\propto v_{rec}\propto v^{-1}$ (see (51) or (54)). This provides a basis for the fundamental comprehension
 of the quantum uncertainties emerging from the space-time of SSR. The non-locality behavior of a ``particle" close to the
 ultra-referential $S_V$ in the space-time of SSR and its implications for QM should be investigated elsewhere.

 It is interesting to observe that we may write $\Delta x^{\prime}_v$ in the following way:

\begin{equation}
\Delta x^{\prime}_v = \frac{(V\Delta\tau)(c\Delta\tau)}{v\Delta\tau}
\equiv\frac{\Delta x^{\prime}_5\Delta x^{\prime}_4}{\Delta x^{\prime}_1},
\end{equation}
where $V\Delta\tau=\Delta x^{\prime}_5$, $c\Delta\tau=\Delta x^{\prime}_4$ and $v\Delta\tau=\Delta x^{\prime}_1$. We know that
 $c\Delta t=\Delta x_4$ and $v\Delta t=\Delta x_1$. So we rewrite (48), as follows:

\begin{equation}
\Delta x^{\prime 2}_4-\frac{\Delta x^{\prime 2}_5\Delta x^{\prime 2}_4}{\Delta x^{\prime 2}_1}=\Delta x_4^2 - \Delta x_1^2
\end{equation}

 If $\Delta x^{\prime}_5\rightarrow 0$ ($V\rightarrow 0$), this implies $\Delta x^{\prime}_v=0$. So we recover the invariance of the
 4-interval in Minkowski space-time, namely $\Delta S^2=(\Delta x_4^2-\Delta x_1^2)=\Delta S^{\prime 2}=\Delta x^{\prime 2}_4$.

 As we have $\Delta x^{\prime}_v>0$, we observe that $\Delta S^{\prime 2}=\Delta x^{\prime 2}_4>\Delta S^2=(\Delta x_4^2-\Delta x_1^2$).
 Thus we may write (59), as follows:

\begin{equation}
\Delta S^{\prime 2} = \Delta S^2 + \Delta x^{\prime 2}_v,
\end{equation}
where $\Delta S^{\prime}=\overline {AC}$, $\Delta x^{\prime}_v=\overline {O^{\prime}C}$ and $\Delta S=\overline {AO^{\prime}}$
(refer to figure 4).

 For $v>>V$ or also $v\rightarrow c$, we have $\Delta S^{\prime}\approx\Delta S$, hence $\theta\approx\frac{\pi}{2}$ (figure 4). In the
 approximation for the macroscopic world (large masses), we have $\Delta x^{\prime}_v=\Delta x^{\prime}_5=0$ (hidden dimension); so
 $\theta=\frac{\pi}{2}\Rightarrow\Delta S^{\prime}=\Delta S$ ($V=0$).

   For $v\rightarrow V$, we would have $\Delta S^{\prime}>>\Delta S$, so that $\Delta S^{\prime}\approx\Delta x^{\prime}_v\approx
 c\Delta\tau$ since $\Delta\tau\rightarrow\infty$ and $\theta\rightarrow\pi$. In this new relativistic limit (ultra-referential $S_V$),
 due to the infinite delocalization $\Delta x^{\prime}_v\rightarrow\infty$, the 4-interval $\Delta S^{\prime}$ loses completely its
 equivalence with respect to $\Delta S$ because we have $\Delta x^{\prime}_5\rightarrow\infty$ (see (59)).

    Equation (60) (or (59)) shows us a break of $4$-interval invariance ($\Delta S^{\prime}\neq\Delta S$), which becomes noticeable
 only in the limit $v\rightarrow V$ (close to $S_V$, i.e., $\Delta x^{\prime}_v\rightarrow\infty$). However, a new invariance is restored
 when we implement an effective intrinsic dimension ($x^{\prime}_5$) for the moving particle at its non-Galilean frame $S^{\prime}$ by
 means of the definition of a new interval, namely:

\begin{equation}
\Delta S_5=\Delta x^{\prime}_4\sqrt{1-\frac{\Delta x^2_5}{\Delta x^{\prime 2}_1}}=c\Delta\tau\sqrt{1-\frac{V^2}{v^2}},
\end{equation}
so that $\Delta S_5\equiv\Delta S=\sqrt{\Delta S^{\prime 2}-\Delta x^{\prime 2}_v}$ (see (60)).

 We have omitted the index $\prime$ of $\Delta x_5$, as such interval is given only at the non-Galilean proper frame ($S^{\prime}$),
being an intrinsic (proper) dimension of the particle. However, from a practical viewpoint, namely for experiences of higher energies, the
electron approximates more and more to a punctual particle since $\Delta x_5$ becomes hidden.

 Actually the new interval $\Delta S_5$, which could be simply denominated as an effective 4-interval
 $\Delta S=c\Delta\tau*=c\Delta\tau\sqrt{1-\alpha^2}$, guarantees the existence of a certain non-null internal dimension of the particle
 (see (61)), which leads to $\Delta x^{\prime}_v>0$ and thus $v_{rec}\neq 0(>V)$.

Comparing (61) with the left-hand side of equation (56), we may alternatively write

\begin{equation}
\Delta t=\Psi\Delta\tau=\frac{\Delta S_5}{c\sqrt{1-\frac{v^2}{c^2}}}=
\Delta\tau\frac{\sqrt{1-\frac{V^2}{v^2}}}{\sqrt{1-\frac{v^2}{c^2}}} ,
\end{equation}
where $\Delta S_5$ corresponds to the invariant effective 4-interval, i.e., $\Delta S_5\equiv\Delta S$
(segment $\overline{AO^{\prime}}$ in figure 4).

 We have $\Psi=\frac{\sqrt{1-\alpha^2}}{\sqrt{1-\beta^2}}=\frac{\sqrt{1-\frac{V^2}{v^2}}}{\sqrt{1-\frac{v^2}{c^2}}}$ and we can
 alternatively write
 $\Psi=\frac{\sqrt{1-\beta^2_{int}}}{\sqrt{1-\alpha^2_{int}}}=\frac{\sqrt{1-\frac{v_{int}^2}{c^2}}}{\sqrt{1-\frac{V^2}{v_{int}^2}}}$,
 since $\alpha=V/v=\beta_{int}=v_{int}/c$ and $\beta=v/c=\alpha_{int}=V/v_{int}$, from where we get $v_{int}=v_{rec}=cV/v=v_0^2/v$
 (see (54)).

  Although we cannot obtain $v_{rec}$ by any direct experience, we could also consider $\Psi$ in its alternative form $\Psi(v_{rec})$.
  However, by convenience, let us simply use $\Psi(v)$ instead of $\Psi(v_{rec})$.

  For $v>>V$ or $V\rightarrow 0$, we get the approximation $\Delta t\approx\gamma\Delta\tau$, where $\Psi\approx\gamma=(1-\beta^2)^{-1/2}$
 (Lorentz factor). In SR theory we have $0<v<c$ and $v_{rec}=0$.

  Inserting (57) into (48), we obtain

\begin{equation}
c^2\Delta\tau^2=\frac{1}{\left(1-\frac{V^2}{v^2}\right)}[c^2\Delta t^2-v^2\Delta t^2],
\end{equation}
from where we also obtain the time equation (56).

 From (63), we can also verify that the proper space-time interval $\Delta S^{\prime}$ ($c\Delta\tau$) dilates drastically in the
limit $v\rightarrow V$, i.e., $\Delta S^{\prime}=c\Delta\tau\rightarrow\infty$. In order to describe such an effect in terms of
metric, we write $dS^{\prime 2}=dS_v^2=\Theta_v dS^2=\Theta_vg_{\mu\nu}dx^{\mu}dx^{\nu}$, where we have
$\Theta_v=\Theta(v)=\left(1-\frac{V^2}{v^2}\right)^{-1}$, being $\Theta(v)$ the dilatation factor\cite{12}, leading to an effective
 (deformed) metric that depends on the speed $v$, namely $G_{\mu\nu}(v)=\Theta(v)g_{\mu\nu}$. So we can write
$dS_v^2=G_{\mu\nu}(v)dx^{\mu}dx^{\nu}=\frac{1}{\left(1-\frac{V^2}{v^2}\right)}[c^2dt^2-dx^2-dy^2-dz^2]$, where
 $\Delta S_v=\Delta S^{\prime}$ (dilated interval). Of course if we make $V\rightarrow 0$ or $v>>V$, there would not be dilatation factor,
 i.e., $\Theta_v\approx 1$, recovering the Minkowski metric. By considering $dy=dz=0$ and $dx=dx_1=vdt$ above, we obtain eq.(63) leading
to the time equation (56).

  As the new metric $G_{\mu\nu}(v)$ of SSR depends on velocity, it seems to be related to a kind of Finsler metric, namely a Finslerian
(non-Riemannian) space with a metric depending on position and velocity, that is, $g_{\mu\nu}(x,\dot{x})$\cite{36}\cite{37}
 \cite{38}\cite{39}\cite{40}. Of course, if there is no dependence on velocity, the Finsler space turns out to be a Riemannian space. Such
 a connection between $G_{\mu\nu}(v)$ and Finslerian geometry should be investigated.

By placing eq.(63) in a differential form and manipulating it, we will obtain

\begin{equation}
c^2\left(1-\frac{V^2}{v^2}\right)\frac{(d\tau)^2}{(dt)^2} + v^2 = c^2
\end{equation}

We may write (64) in the following alternative way:

\begin{equation}
\frac{(dS_5)^2}{(dt)^2} + v^2 = c^2,
\end{equation}
where $dS_5=c\sqrt{1-\frac{V^2}{v^2}}d\tau$.

 Equation (64) shows us that the speed related to the marching of the time (``temporal-speed''), that is
 $v_t=c\sqrt{1-\frac{V^2}{v^2}}\frac{d\tau}{dt}$, and the spatial speed $v$ with respect to the background field ($S_V$) form the
legs of a rectangle triangle according to figure 5.

\begin{figure}
\includegraphics[scale=0.7]{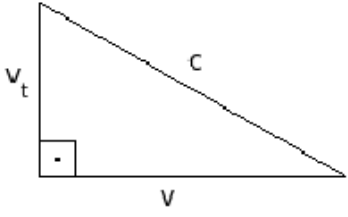}
\caption{We have $c=(v_t^2+v^2)^{1/2}$, which represents the space-temporal speed of any particle (hypothenuse of the triangle=$c$).
 The novelty here is that such a structure of space-time implements the ultra-referential $S_V$. This implementation arises at the vertical
 leg $v_t$.}
\end{figure}

  In accordance with figure 5, we should consider 3 importants cases, namely:

 a)    If $v\approx c$, then $v_t\approx 0$, where $\Psi>>1$, since $\Delta t>>\Delta\tau$ ({\it dilatation of improper time}).

 b)    If $v=v_0$, then $v_t=\sqrt{c^2-v_0^2}$, where $\Psi=\Psi_0=\Psi(v_0)=1$, since $\Delta t=\Delta\tau$ ({\it ``quantum rest''}).

 c)    If $v\approx V$, then $v_t\approx\sqrt{c^2-V^2} =c\sqrt{1-\xi^2}$, where $\Psi<<1$, since $\Delta t<<\Delta\tau$ ({\it contraction
 of improper time}).

  In SR, for $v=0$, we have $v_t=v_{tmax}=c$. However, in accordance with SSR, due to the existence of a minimum limit of spatial speed $V$
 for the horizontal leg of the triangle, we realize that the maximum temporal speed is $v_{tmax}<c$, namely we have
 $v_{tmax}=c\sqrt{1-\xi^2}$. Such a result introduces a strong symmetry for SSR in the sense that both spatial and temporal speeds $c$
 become forbidden for all massive particles.

 The speed $v=c$ is represented by the photon (particle without mass), whereas $v=V$ is definitely forbidden for any particle. So
 we generally have $V<v\leq c$. But, in this sense, we have a certain asymmetry as there is no particle at the ultra-referential $S_V$
 where there should be a kind of {\it sui generis} vacuum energy\cite{12}.

  In order to produce a geometric representation for this problem ($V<v\leq c$), let us assume the world line of a particle limited by the
 surfaces of two cones (figure 6).

\begin{figure}
\begin{center}
\includegraphics[scale=0.6]{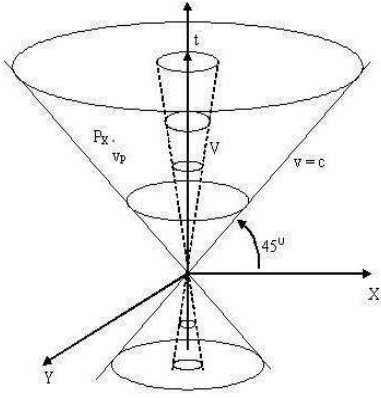}
\end{center}
\caption{The external and internal conical surfaces represent respectively $c$ and $V$, where $V$ is represented by the dashed line, namely
 a definitely prohibited boundary. For a point $P$ in the interior of the two conical surfaces, there is a corresponding internal conical
 surface so that $V<v_p\leq c$.}
\end{figure}

 The spatial speed ($v_p$) in the representation of light cone given in figure 6 (the horizontal leg of the triangle in figure 5)
 is associated with a temporal speed $v_{tp}=\sqrt{c^2-v_p^2}$ (the vertical leg of the same triangle) given in another cone
 representation, which could be denominated as {\it temporal cone} (see figure 7).

\begin{figure}
\begin{center}
\includegraphics[scale=0.6]{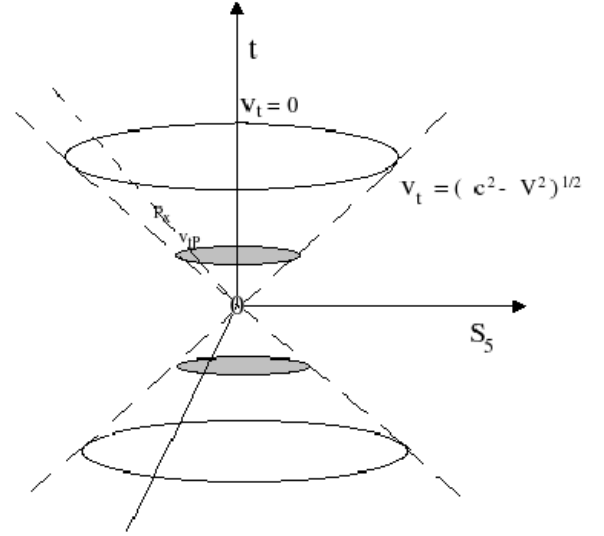}
\end{center}
\caption{Comparing this figure with figure 6, we notice that the dashed line on the internal cone of figure 6 ($v=V$) corresponds to
 the dashed line on the surface of the external cone of this figure, where $v_t=\sqrt{c^2-V^2}$, which represents a definitely forbidden
 boundary in this cone representation of temporal speed $v_t$. On the other hand, $v=c$ (photon) is represented by the solid line of figure
 6, which corresponds to the temporal speed $v_t=0$ in this figure. In short, we always have $v^2+v_t^2=c^2$, being $v$ for the spatial
 (light) cone (figure 6) and $v_t$ for the temporal cone represented in this figure 7.}
\end{figure}

   In figure 6, we see two boundary surfaces ($c$ and $V$), whereas, in figure 7, we observe just one external
 boundary surface ($\sqrt{c^2-V^2}$). Such a difference reflects a certain fundamental asymmetry which does not occur in SR where there is
 one single external boundary cone (solid line) in the spatial representation ($0\leq v_p\leq c$) as well as in the temporal one
($0\leq v_{tp}\leq c$).

 Based on equation (63) or also by inserting (57) into (48), we obtain

  \begin{equation}
  c^2\Delta t^2 - v^2\Delta t^2 = c^2\Delta\tau^2 -
 \frac{v_0^4}{v^2}\Delta\tau^2
  \end{equation}

 In eq.(66), when we transpose the 2nd term from the left-hand side to the right-hand side and divide the equation by $(\Delta t)^2$, we
 obtain (64) given in differential form. Now, it is important to observe that, upon transposing the 2nd term from the right-hand side to
 the left-hand one and dividing the equation by $(\Delta\tau)^2$, we obtain the following equation in differential form, namely:

  \begin{equation}
  c^2\left(1-\frac{v^2}{c^2}\right)\frac{(dt)^2}{(d\tau)^2} + \frac{v_0^4}{v^2}=c^2
  \end{equation}

 From (61) and (56), we obtain $dS_5=cd\tau\sqrt{1-\alpha^2}=cdt\sqrt{1-\beta^2}$. Hence we can write (67) in the following alternative
 way:

  \begin{equation}
  \frac{(dS_5)^2}{(d\tau)^2} + \frac{v_0^4}{v^2}= c^2
  \end{equation}

  Equation (67) (or (68)) reveals a complementary way of viewing equation (64) (or (65)), which takes us to that idea of
 reciprocal space for conjugate quantities. Thus let us write (67) (or (68)) in the following way:

  \begin{equation}
  v_{trec}^2 + v_{rec}^2 =c^2,
  \end{equation}
where $v_{trec}(=(v_t)_{int}=c\sqrt{1-\frac{v^2}{c^2}}\frac{dt}{d\tau})$ represents an internal (reciprocal)``temporal speed"
 and $v_{rec}(=v_{int}=f(v)=\frac{v_0^2}{v})$ is the internal (reciprocal) spatial speed. Therefore we can represent a rectangle
 triangle which is similar to that of figure 5, but now being represented in a reciprocal space. For example, if we assume $v\rightarrow c$
 (equation (64)),
we obtain $v_{rec}=lim._{v\rightarrow c}f(v)\rightarrow V$ (equation (67)). For this same case, we have $v_t\rightarrow 0$ (equation (64))
 and $v_{trec}=\frac{dS_5}{d\tau}\rightarrow\sqrt{c^2-V^2}$ (equation (67) or (68)). On the other hand, if $v\rightarrow V$ (eq.(64)), we
 have $v_{rec}\rightarrow\frac{v_0^2}{V}=c$ (eq.(67)), where $v_t\rightarrow\sqrt{c^2-V^2}$ (eq.(64)) and $(v_t)_{int}=v_{trec}\rightarrow
 0$ (eq.(67)). Thus we should observe that there are altogheter four cone representations in SSR, namely:

\begin{equation}
 spatial~representations:\left\{\begin{array}{ll}
a_1)  v=\frac{dx}{dt}, in~eq.(64),\\
represented~in~Fig.6;\\
 b_1) v_{rec}=\frac{d x^{\prime}_v}{d\tau}=\frac{v_0^2}{v},\\
 in~eq.(67).
\end{array}
 \right.
\end{equation}

\begin{equation}
 temporal~representations:\left\{\begin{array}{ll}
a_2) v_t=c\sqrt{1-\frac{V^2}{v^2}}\frac{d\tau}{dt}\\
=c\sqrt{1-\frac{v^2}{c^2}}, in~eq.(64),\\
represented~in~Fig.7;\\

b_2) v_{trec}=c\sqrt{1-\frac{v^2}{c^2}}\frac{dt}{d\tau}\\
=c\sqrt{1-\frac{V^2}{v^2}}, in~eq.(67).
\end{array}
 \right.
 \end{equation}

The chart in figure 8 shows us the four representations.

\begin{figure}
\begin{center}
\includegraphics[scale=0.55]{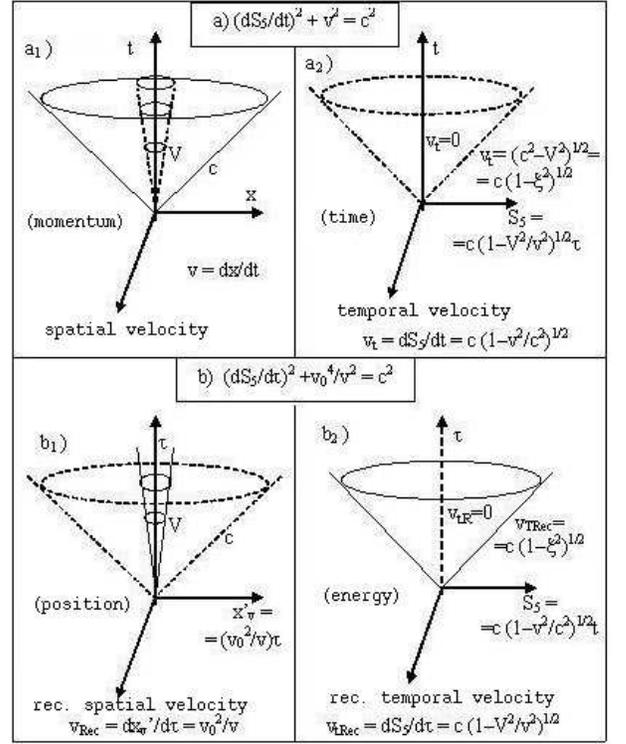}
\end{center}
\caption{The spatial representations in $a_1$ (also shown in figure 6) and $b_1$ are related respectively to velocity $v$ (momentum) and
 position (delocalization $\Delta x^{\prime}_v=f(v)\Delta\tau=v_{int}\Delta\tau=v_{rec}\Delta\tau=(v_0^2/v)\Delta\tau$), which
 represent conjugate (reciprocal) quantities in the space. On the other hand, the temporal representations in $a_2$ (also shown in figure
 7) and $b_2$ are related respectively to the time ($\propto v_t$) and the energy ($\propto v_{trec}=(v_t)_{int}\propto v_t^{-1}$), which
 represent (reciprocal) conjugate quantities in the time. Hence we can perceive that such four cone representations of SSR provide a basis
 for the fundamental understanding of the two uncertainty relations.}
\end{figure}

Now, by considering (56),(62) and (71), looking at $a_2$ and $b_2$ in figure 8, we may observe

 \begin{equation}
 \Psi^{-1}=\frac{\Delta\tau}{\Delta t}= \frac{\sqrt{1-\frac{v^2}{c^2}}}{\sqrt{1-\frac{V^2}{v^2}}}=
\frac{v_t}{c\sqrt{1-\frac{V^2}{v^2}}}= \frac{v_t}{v_{trec}}\propto (time)
  \end{equation}

  and

  \begin{equation}
\Psi=\frac{\Delta t}{\Delta\tau}= \frac{\sqrt{1-\frac{V^2}{v^2}}}{\sqrt{1-\frac{v^2}{c^2}}}=
\frac{v_{trec}}{c\sqrt{1-\frac{v^2}{c^2}}}=\frac{v_{trec}}{v_t}\propto E,
  \end{equation}
being $E=Energy\propto (time)^{-1}$.

 From (73), as we have $E\propto\Psi$, we obtain $E=E_0\Psi$, where $E_0$ is a constant. Hence, by considering
 $E_0=m_0c^2$, we write

\begin{equation}
E= m_0c^2\frac{\sqrt{1-\frac{V^2}{v^2}}}{\sqrt{1-\frac{v^2}{c^2}}},
\end{equation}
where $E$ is the total energy of the particle with respect to the ultra-referential $S_V$ of the background field. In (73) and (74), we
 observe that, if $v\rightarrow c\Rightarrow E\rightarrow\infty$ and $\Delta\tau\rightarrow 0$ for $\Delta t$ fixed; if $v\rightarrow
 V\Rightarrow E\rightarrow 0$ and $\Delta\tau\rightarrow\infty$, also for $\Delta t$ fixed. If $v=v_0=\sqrt{cV}\cong 5.65\times 10^{-4}
m/s\Rightarrow E=E_0=m_0c^2$ (energy of ``quantum rest'' at $S_0$).

  Figure 9 shows us the graph for the energy $E$ in eq.(74).

\begin{figure}
\begin{center}
\includegraphics[scale=0.6]{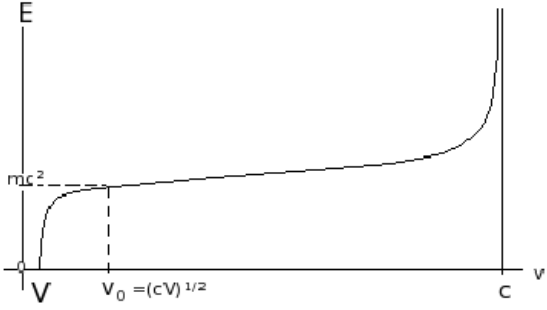}
\end{center}
\caption{$v_0$ represents the speed in relation to $S_V$, from where we get the proper energy of the particle ($E_0=m_0c^2$), being
 $\Psi_0=\Psi(v_0)=1$. For $v<<v_0$ or closer to $S_V$ ($v\rightarrow V$), a new relativistic correction on energy arises, so that
 $E\rightarrow 0$.}
\end{figure}

\subsection{Deformed relativistic dynamics in SSR}

Let us introduce some aspects of the deformed relativistic dynamics in SSR\cite{12}. So we firstly define the 4-velocity in the presence
of $S_V$, as follows:

\begin{equation}
 U^{\mu}=\left[\frac{\sqrt{1-\frac{V^2}{v^2}}}{\sqrt{1-\frac{v^2}{c^2}}}~ , ~
\frac{v_{\alpha}\sqrt{1-\frac{V^2}{v^2}}}{c\sqrt{1-\frac{v^2}{c^2}}}\right],
\end{equation}
where $\mu=0,1,2,3$ and $\alpha=1,2,3$. If $V\rightarrow 0$, we recover the well-known 4-velocity of SR. From (75) it is interesting to
 observe that the 4-velocity of SSR vanishes in the limit of $v\rightarrow V$ ($S_V$), i.e., $U^{\mu}=(0,0,0,0)$, whereas for $v=0$ in
SR, we find $U^{\mu}=(1,0,0,0)$.

 The 4-momentum is
\begin{equation}
 P^{\mu}=m_0cU^{\mu},
   \end{equation}
being $U^{\mu}$ given in (75). So we find

\begin{equation}
 P^{\mu}=\left[\frac{m_0c\sqrt{1-\frac{V^2}{v^2}}}{\sqrt{1-\frac{v^2}{c^2}}}~ , ~
\frac{m_0v_{\alpha}\sqrt{1-\frac{V^2}{v^2}}}{\sqrt{1-\frac{v^2}{c^2}}}\right],
\end{equation}
where $P^0=E/c$, such that

\begin{equation}
E=cP^0=mc^2=m_0c^2\frac{\sqrt{1-\frac{V^2}{v^2}}}{\sqrt{1-\frac{v^2}{c^2}}},
\end{equation}
where $E$ is the total energy of the particle (figure 9).

From (77) we also obtain the momentum with respect to $S_V$, namely:

\begin{equation}
\vec P = m_0\vec v\frac{\sqrt{1-\frac{V^2}{v^2}}}{\sqrt{1-\frac{v^2}{c^2}}},
\end{equation}
where $P_{\alpha}$ ($\alpha=1,2,3$) are the spatial components of the 4-momentum $P^{\mu}$.

For $v=v_0$, we find $P=m_0v_0=m_0\sqrt{cV}$, as $\Psi(v_0)=1$.

From (77), by performing the quantity $P^{\mu}P_{\mu}$, we obtain the energy-momentum relation of SSR, as follows:

\begin{equation}
P^{\mu}P_{\mu}=\frac{E^2}{c^2}-\vec P^2=m_0^2c^2\left(1-\frac{V^2}{v^2}\right)
\end{equation}

 From (80) we obtain

\begin{equation}
E^2=c^2P^2+m_0^2c^4\left(1-\frac{V^2}{v^2}\right)
\end{equation}

 If we make $V\rightarrow 0$ above, we recover the well-known energy-momentum relation of SR.

 Others aspects of such a deformed relativistic dynamics as well as the deformed algebras in SSR should be investigated elsewhere.

 \section{\label{sec:level1} The origin of the uncertainty principle}

 \subsection{The intrinsic uncertainty of a particle in the space-time of SSR}

  The particle actual momentum with respect to $S_V$ is $P=\Psi m_0 v$, whose conjugate value is $\Delta x^{\prime}_v=v_{rec}\Delta\tau=
 \frac{v_0^2}{v}\Delta\tau=\frac{v_0^2}{v}\Delta t\Psi^{-1}$, where $\Delta\tau=\Psi^{-1}\Delta t$ (refer to (56)). Since
  $\Delta x^{\prime}_v$ represents an ``internal displacement" (delocalization) working like an intrinsic uncertainty on position of the
 particle, the momentum $P$ which represents its conjugate value should be also interpreted as an intrinsic uncertainty on
 momentum, namely $P=(\Delta p)_0$.

  As $P$ is the actual momentum given with respect to the ultra-referential $S_V$, it is always inaccessible
 to a classical observer at any Galilean reference frame $S$ at rest. Due to this impossibility to know exactly the actual momentum $P$
 from any Galilean frame $S$, so $P$ appears as an intrinsic uncertainty $(\Delta p)_0$ on the wave-packet of the particle. In other words
 this means that the speed $v$ (non-Galilean frame $S^{\prime}$) given with respect to $S_V$ appears as $(\Delta v)_0$ at any inertial
 (Galilean) reference frame $S$ at rest, i.e., $v=(\Delta v)_0$. For example, if $v\rightarrow V$ ($P\rightarrow 0$), this means
 $(\Delta v)_0\rightarrow V$ ($(\Delta p)_0\rightarrow 0$), and so we have $\Delta x^{\prime}_v\rightarrow\infty$. In this case we get
 the wave-function of the particle close to a plane wave, but never an exact one since $V$ is unattainable.

  We must stress that the intrinsic aspect of the uncertainties $\Delta x^{\prime}_v$ and $(\Delta p)_0$ has purely origin from the nature
 of the space-time in SSR (see figure 8), where the classical observer is not taken under consideration just in the sense that he does not
 attempt to measure the momentum and position of the moving particle (electron). However, if we want to derive the well-known uncertainty
 principle as given in QM, we have to take in account the classical observer who tries to measure the momentum and position of the particle
 by emitting a photon that interacts with it. But before to do that, let us firstly obtain the intrinsic uncertainty ($I_{0}$) emerging
 naturally from the space-time of SSR, namely:

\begin{equation}
I_{0}=\Delta x^{\prime}_vP=\frac{v_0^2}{v}\Delta t\Psi^{-1}\Psi m_0 v=(m_0v_0)(v_0\Delta t),
\end{equation}
 where $P=(\Delta p)_{0}$ (no observer). In obtaining (82), we have also considered the relations
 $\Delta x^{\prime}_v=\frac{v_0^2}{v}\Delta\tau$, $\Delta\tau=\Delta t\Psi^{-1}$ and $(\Delta p)_0=P=\Psi m_0 v$. We have
 $v_0=\sqrt{cV}$. Of course we get $I_{0}=0$ in SR since $V=0$ ($v_0=0$).

  The fundamental reason why the actual speed $v$ (with respect to $S_V$) works like the width $(\Delta v)_{0}$ on the wave-packet
of the particle at any Galilean (inertial) frame $S$ has origin from the nature of the non-Galilean frames in SSR compared with the
well-known inertial nature of the Galilean reference frames in SR.

  We have defined a non-Galilean frame $S^{\prime}$ as a set of all the particles with a speed $v$ with respect to $S_V$, however
such a frame $S^{\prime}$ cannot be understood as subjected to the notion of relative rest (equivalence of inertial systems). In this
sense, for instance, two non-Galilean frames $v$ ($S^{\prime}$) and $v^{\prime}$ ($S^{\prime\prime}$) are not equivalent to each other
 and thus they work effectively like non-inertial frames whose states of motion should be absolute, namely $v$ and $v^{\prime}$ are
 absolute speeds with respect to the preferred non-Galilean background frame ($S_V$). In short, in SSR we say that the presence of the
 preferred frame $S_V$ of background field breaks the equivalence of the reference frames since they behave effectively
 like non-inertial frames (non-Galilean frames) having their absolute states of motion as, for instance, the speed $v$ of $S^{\prime}$ with
 respect to $S_V$.

 As a non-inertial frame has the same state of motion (e.g: acceleration) given for any inertial (Galilean) frame, we also expect that
 a non-Galilean frame of SSR, having an absolute speed $v$ ($V<v<c$) with respect to the preferred frame $S_V$, remains invariant for any
 inertial (Galilean) frame. However, since the absolute speed $v$ is inaccessible for us at any Galilean (inertial) reference frame, it
 appears as an intrinsic width $(\Delta v)_0$ on the wave-packet of the moving particle in order to reveal us the non-inertial aspect of
 its wave-function.

 The non-inertial aspect on the wave-function of the moving particle has to do with the fact that there is no free particle in Nature 
 since the ideal case of a free particle (a plane wave wave-function) corresponds effectively to an inertial system where the width of the
 wave packet would be $(\Delta v)_{0}=0$. And as we should have $(\Delta v)_{0}>V$, indeed there is no free particle and so the presence
of the non-inertial aspect on the wave-function cannot be completely eliminated. Hence we conclude that the non-inertial aspect (width
$(\Delta v)_{0}$) on the wave-packet of the particle moving with respect to any Galilean frame $S$ has origin from its non-Galilean frame
with speed $v$ with respect to $S_V$ since we have $(\Delta v)_{0}=v$. So $v$ appears as $(\Delta v)_{0}$ for any Galilean frame since $v$
($S^{\prime}$) is non-Galilean.

  If $V\rightarrow 0$ (or $G\rightarrow 0$ in (43)), the non-inertial aspect of the wave-function would vanish ($(\Delta v)_{0}=0$) and
so we would have the ideal case of a free particle (a plane wave wave-function) having an inertial behavior. Of course this would happen
only in the absence of interactions over the particle, specially the hypothetical absence of gravity ($G=0$). However the complete absence
of gravity is not possible since its source could be any mass (or energy), and besides this it has an infinite range in the space. As the
minimum speed $V$ is directly connected to gravity, i.e., $V\propto\sqrt{G}$ (see (43) or (44)), we realize that the non-inertial aspect on
the wave-packet of the particle has essentially origin from gravity ($(\Delta v)_{0}>V\propto\sqrt{G}$), leading to the uncertainty
principle. So we conclude that the minimum speed $V$ connected to the background frame $S_V$ provides a basis for
understanding new aspects of a quantum gravity theory at low energies, from where the quantum uncertainties naturally emerge.

 \subsection{The uncertainty principle}

 Even when there is no observer to detect the uncertainty on momentum and position of the particle, we have $I_{0}$ as an intrinsic
uncertainty emerging from the space-time of SSR. However, the presence of a classical observer who tries to measure its momentum and
 position by emitting a photon towards to the particle leads to a perturbation on $I_{0}$, so that we have

 \begin{equation}
 I=I_{0}+\delta I_{0},
 \end{equation}
where ``$I$" should be the well-known uncertainty of QM. $\delta I_{0}$ is a variation on the intrinsic uncertainty $I_{0}$ due to the
scattering of the photon on the particle (electron).

  In order to compute the perturbation $\delta I_{0}$ on the intrinsic uncertainty $I_0$ of the electron, we should realize that its
 intrinsic uncertainties $P$($=(\Delta p)_0$) and $\Delta x^{\prime}_v$ suffer variations (perturbations) due to its interaction with the
 photon, so that we have

 \begin{equation}
 \delta I_{0}=\delta(\Delta x^{\prime}_v P)=\Delta x^{\prime}_v(\delta P) + P\delta(\Delta x^{\prime}_v)
 \end{equation}
from where we get $\delta I_{0}=\Delta x^{\prime}_v\delta p+P\delta x$, where $\delta P\equiv\delta p$ and
$\delta(\Delta x^{\prime}_v)\equiv\delta x$. The quantities $\delta p$ and $\delta x$ represent respectively the variations on the
intrinsic uncertainties of momentum and position of the electron due to the photon scattering.

Let us compute each term of $\delta I_{0}$, as follows:

a)1st.term: $\Delta x^{\prime}_v\delta p=\frac{v_0^2}{v}\Delta\tau\delta p=\frac{v_0^2}{v}\Psi^{-1}\Delta t\delta p$. In order to compute
$\delta p$, we should consider the Compton scattering such that a photon is emitted towards to the particle (electron) across the path
of length $L$. The scattered photon returns in the same path $L$ towards to the observer. So the interval of time elapsed during the
 emission and detection of the photon is $\Delta t=\frac{2L}{c}$. In this case, we have the scattered angle $\Phi=\pi$. Thus, the deviation
 of wavelength of the emitted photon is maximum, i.e., $\Delta\lambda_{max}=\lambda-\lambda_{0}=
\frac{c}{\nu}-\frac{c}{\nu_{0}}=\frac{2h}{m_0c}$, where $\lambda_0$ and $\nu_0$ are respectively the wave-length and the frequency of
the emitted photon. The quantities $\lambda$ and $\nu$ are associated with the scattered photon.

  Hence we find $\delta p=p-p_{0}=\frac{h(\nu_{0}-\nu)}{c}=\frac{2h^2\nu\nu_0}{m_0c^3}$, being $\delta p$ the momentum transferred to the
 electron. Finally we obtain the first term, as follows:

\begin{equation}
 \Delta x^{\prime}_v\delta p=\frac{v_0^2}{v}\frac{\sqrt{1-\frac{v^2}{c^2}}}{\sqrt{1-\frac{V^2}{v^2}}}
\left(\frac{2h^2\nu\nu_0}{m_0c^3}\right)\left(\frac{2L}{c}\right),
 \end{equation}
where $v_0^2=cV$.

 In non-relativistic (newtonian) approximation for SSR, where we consider $V<<v<<c$, from (85) we get

\begin{equation}
 \Delta x^{\prime}_v\delta p\approx\frac{v_0^2}{v}\left(\frac{2h^2\nu\nu_0}{m_0c^3}\right)\left(\frac{2L}{c}\right),
 \end{equation}
where $\Psi\approx 1$. We observe $v_{rec}=\frac{v_0^2}{v}$ above in (86).

b)2nd.term: $P\delta x=m_0v\Psi\delta x$. Since we also consider the case of $\Psi\approx 1$, we obtain $P\delta x\approx m_0v\delta x$.
 We have $\delta x=\delta v\Delta t=\frac{\delta p}{m_0}\frac{2L}{c}=\left(\frac{2h^2\nu\nu_0}{m_0^2c^3}\right)\left(\frac{2L}{c}\right)$.
 So we finally get

\begin{equation}
 P\delta x\approx v\left(\frac{2h^2\nu\nu_0}{m_0c^3}\right)\left(\frac{2L}{c}\right)
 \end{equation}

 We notice that the two terms (86) and (87) represent reciprocal quantities to each other in the space-time of SSR since
 $v_{rec}=\frac{v_0^2}{v}$ (see (86)).

 From (84), (86) and (87), we finally obtain

\begin{equation}
 \delta I_{0}=\left[v+\frac{v_0^2}{v}\right]\left(\frac{2h^2\nu\nu_0}{m_0c^3}\right)\left(\frac{2L}{c}\right)
 \end{equation}

 From (82) we write

\begin{equation}
 I_{0}= m_0v_0^2\left(\frac{2L}{c}\right),
 \end{equation}
where $\Delta t=2L/c$.

 We have $I=I_{0}+\delta I_{0}=\Delta x^{\prime}_v P+\delta(\Delta x^{\prime}_v P)=
\Delta x^{\prime}_v P+\Delta x^{\prime}_v(\delta P) + P\delta(\Delta x^{\prime}_v)$ (see (82), (83) and (84)).

 According to (88) we see that $\delta I_{0}$ is a function of $v$($=(\Delta v)_0$), having a minimum value $(\delta I_{0})_{min}$
for a certain value of $v$. Such a minimum value leads to a minimum value of uncertainty $I_{min}$, namely
 $I_{min}=I_{0}+(\delta I_{0})_{min}$. In order to compute $(\delta I_{0})_{min}$, first of all we must obtain the value of $v$ that
 minimizes $\delta I_{0}$, as follows:

\begin{equation}
\frac{d(\delta I_{0})}{dv}=\left[1-\frac{v_0^2}{v^2}\right]\left(\frac{2h^2\nu\nu_0}{m_0c^3}\right)\left(\frac{2L}{c}\right)=0,
 \end{equation}
which implies $v=v_0$. So by inserting this point of minimum $v_0$ into (88), we obtain

\begin{equation}
 (\delta I_{0})_{min}=\left[2 v_0\right]\left(\frac{2h^2\nu\nu_0}{m_0c^3}\right)\left(\frac{2L}{c}\right)
 \end{equation}

 It is interesting to notice that, just at the point of minimum ($v_0$), the mixing terms (86) and (87) are equal to each other.
 So they just contribute equally for $(\delta I_{0})_{min}$, namely $\Delta x^{\prime}_v\delta p]_{v=v_0}= P\delta x]_{v=v_0}=
v_0\left(\frac{2h^2\nu\nu_0}{m_0c^3}\right)\left(\frac{2L}{c}\right)$, being $v_0=\sqrt{cV}$. Of course we verify that, if
$V\rightarrow 0$, those two terms vanish and so we recover the classical space-time of SR where there are no uncertainties.

  In order to estimate the magnitude of $(\delta I_{0})_{min}$, we use a $\gamma$-ray that is going to impact the electron according to the
experiment of Heisenberg's microscope. The emitted $\gamma$-photon with $\nu_0\sim 10^{19}$Hz is scattered by the electron. As the
scattered photon comes back to the observer in the same path of length $L$, it is well-known that the deviation of its wave-length is
maximum according to Compton scattering, namely $(\Delta\lambda)_{max.}=\lambda-\lambda_0\approx 0.049 A^{o}$. The
$\gamma$-ray wave-length is $\lambda_0\sim 10^{-11}$m (the incident photon). So we find the wave-length of the scattered photon, namely
 $\lambda=\lambda_0+(\Delta\lambda)_{max}\sim 10^{-11}$m, being $\nu_{0}\approx\nu\sim 10^{19}$Hz (frequency of the scattered photon).

  Since the scale of length of the classical (human) observer is of the order of $10^{0}$m, the path length $L$ of the photon has
the same order of magnitude, i.e., $L\sim 10^{0}$m. And as we already know $m_0(\sim 10^{-30}Kg$) and $v_0=\sqrt{cV}\sim 10^{-3}$m/s, being
$V\sim 10^{-15}$m/s and $c^4\sim 10^{34}m^4/s^4$, we finally can compute the order of magnitude of $I_{min}$ according to (92), namely
$O(I_{min})=O(I_{0})+O[(\delta I_{0})_{min}]$ to be determined.

 So by considering (89) and (91), we obtain the minimum uncertainty ($I_{min}$), as follows:

\begin{equation}
 I_{min}=\left[m_0v_0^2+2v_0\left(\frac{2h^2\nu\nu_0}{m_0c^3}\right)\right]\left(\frac{2L}{c}\right),
 \end{equation}
where $I_{min}=I_0+(\delta I_0)_{min}$.

  We can alternatively write (92), as follows:

\begin{equation}
 I_{min}=\left[(2m_0V)L+(8m_0\sqrt{cV})\left(\frac{h^2\nu\nu_0}{m_0^2c^4}\right)L\right],
 \end{equation}
from where we get $I_{0}=2m_0VL\sim 10^{-45}$J.s and $(\delta I_{0})_{min}=8m_0\sqrt{cV}\left(\frac{h^2\nu\nu_0}{m_0^2c^4}\right)L\sim
 10^{-34}$J.s. Finally we compute $O(I_{min})=O(I_{0})+O[(\delta I_{0})_{min}]=$ $10^{-45}$J.s + $10^{-34}$J.s $\approx
 10^{-34}$J.s $\sim\hbar$. This result is exactly in agreement with the minimum uncertainty in QM, i.e.,
 $(\Delta x\Delta p)_{min}\sim\hbar$. Of course, if we make $V=0$ in (93), we recover the classical result, i.e., $I_{min}=0$. So we
realize that the non-null minimum speed $V$ in the space-time of SSR provides a fundamental understanding about the origin of the
uncertainty principle. In this sense SSR is consistent with QM.

  When computing $I_{min}$, we have observed that $(\delta I_{0})_{min}>>I_{0}$ so that $I_{min}\approx (\delta I_{0})_{min}\sim\hbar$.
 This means that the uncertainty as known in QM emerges practically from the two mixing terms (86) and (87) given in (88), whose minimun
 value is given in (91). Such mixing terms, where the interaction with the photon appears, play an important role for obtaining the
 uncertainty principle due to the presence of the classical observer. In other words this means that the intrinsic uncertainty $I_{0}$
 (without classical observer) is neglectable when compared with $\delta I_{0}$ obtained in the presence of a classical observer.

  To resume, when $v=v_0$, we have shown that $\delta I_{0}$ assumes a minimum value in the order of $\hbar$, i.e.,
$I_{min}\approx (\delta I_{0})_{min}\sim\hbar$. Hence, for $v>v_0$ or $v<v_0$ (see (88)) we will obtain $I\approx\delta I_{0}>\hbar$
due to the deviation from the minimum point ($v=v_0$). So, in general form, we find

\begin{equation}
I\approx\delta I_{0}=\left[v+\frac{v_0^2}{v}\right]\left(\frac{2h^2\nu\nu_0}{m_0c^3}\right)\left(\frac{2L}{c}\right)\geq\hbar,
\end{equation}
so that, for $v=v_0$, we obtain its minimum value ($\sim\hbar$) and, for $v\neq v_0$, its value increases ($>\hbar$). Such inequality
relation (94) is consistent with the uncertainty relations of QM, i.e., $I=\Delta x\Delta p$ $(\Delta t\Delta E)\geq\hbar$.
 However, we must stress that our result emerges from a fundamental viewpoint of the symmetrical space-time in SSR ($V<v\leq c$).

 The inequality relation (94) is the sum of those two terms given in (86) and (87). As we have shown, for $v=v_0$, both terms contribute
equally for obtaining the uncertainty assuming a minimum value ($\sim\hbar$). But, for $v>v_0$, the term (87) overcomes the term (86),
leading to a deviation from the minimum value of uncertainty. And now, for $v<v_0$, the term (86) overcomes the term (87), also leading
to an increasing of uncertainty ($>\hbar$).

 If $v<<v_0$ (or the limit $v\rightarrow V(S_V)$), we expect $I>>\hbar$ since the term (85) diverges.

\section{\label{sec:level1} Transformations on the fields $\delta\vec E$ and $\delta\vec B$ depending on speed}

 The shift (tiny increment) $\delta\vec E$ (or $\delta\vec B$) has the same direction of $\vec E$ (or $\vec B$) of the electric charge
moving in a constant gravitational potential $\phi$, i.e., we have $\left|\vec E +\delta\vec E(\phi)\right|=\left|\vec E\right|+
\left|\delta\vec E(\phi)\right|=E+\delta E(\phi)$, leading to an increasing of the electromagnetic energy density due to presence of gravity
as shown in eqs.(32) and (33).

 The magnitude of $\delta\vec E$ (or $\delta\vec B$) is given by eqs.(31), namely $\delta E=\xi e_{s0}(\sqrt{\sqrt g_{00}}-1)$ or
$\delta B=\xi b_{s0}(\sqrt{\sqrt g_{00}}-1)$, where $\delta E=\delta E(\phi)$ ($\delta B=\delta B(\phi)$), being $e_{s0}$ and
$b_{s0}$ the magnitudes of electromagnetic fields which are responsible for the proper electromagnetic mass (proper inertial mass) of
the particle, i.e., $m_{0}\propto e_{s0}b_{s0}$.

 According to (18), we get the total energy of the particle having electromagnetic origin, namely
 $E=mc^2\equiv m_{electromag}c^2\propto e_sb_s$ depending on speed $v$ as we have already found in eq.(74), that is,
  $E=mc^2=m_0c^2\Psi(v)$. So comparing (18) with (74), we get

 \begin{equation}
 E=m_0c^2\Psi(v)\propto e_sb_s=e_{s0}b_{s0}\Psi(v),
 \end{equation}
where $m_0c^2\propto e_{s0}b_{s0}$.

 From (95), as the total scalar fields $e_s(v)$ and $b_s(v)$ contribute equally for the total energy $E$ of the particle, we extract
separately the following corrections:

\begin{equation}
  e_s=e_s(v)=e_{s0}\sqrt{\Psi(v)},~ ~
  b_s=b_s(v)=b_{s0}\sqrt{\Psi(v)}
 \end{equation}

 As eqs.(31) were obtained starting from the case of a non-relativistic particle in a constant gravitational potential $\phi$ (see (25)),
 where $K<<m_0c^2$, or even for $V<<v<<c$ according to the newtonian approximation from SSR ($\Psi\approx 1$), so we expect that, for the
case of relativistic corrections with speed in SSR (eqs.(96)), we should make the following corrections in the magnitudes of the fields
given in (31), namely:

\begin{equation}
\delta E(\phi, v)=\xi e_s(\sqrt{\sqrt g_{00}}-1)=\xi e_{s0}\sqrt{\Psi}(\sqrt{\sqrt g_{00}}-1)
\end{equation}

and

\begin{equation}
\delta B(\phi, v)=\xi b_s(\sqrt{\sqrt g_{00}}-1)=\xi b_{s0}\sqrt{\Psi}(\sqrt{\sqrt g_{00}}-1),
\end{equation}
where $\xi=\frac{V}{c}$ and $\sqrt{\Psi}=\sqrt{\Psi(v)}=\frac{\sqrt{\sqrt{1-\frac{V^2}{v^2}}}}{\sqrt{\sqrt{1-\frac{v^2}{c^2}}}}$.

 For $v=v_0=\sqrt{cV}\Rightarrow\Psi(v_0)=1$, so we find $\delta E(\phi,v_0)=
\delta E(\phi)=\xi e_{s0}(\sqrt{\sqrt g_{00}}-1)$ and $\delta B(\phi,v_0)=\delta B(\phi)=\xi b_{s0}(\sqrt{\sqrt g_{00}}-1)$ recovering
eqs.(31). Or even if we make $V<<v<<c$ ($\Psi\approx 1$), we find the newtonian approximation from SSR, where
 $\delta E(\phi,v)\approx\delta E(\phi)$ and $\delta B(\phi,v)\approx\delta B(\phi)$, also recovering the validity of eqs.(31).

 Eqs.(97) and (98) are alternatively written, as follows:

\begin{equation}
\delta E^{\prime}=\frac{\sqrt{\sqrt{1-\frac{V^2}{v^2}}}}{\sqrt{\sqrt{1-\frac{v^2}{c^2}}}}\delta E,
\end{equation}
where $\delta E=\delta E(\phi)$ and $\delta E^{\prime}=\delta E(\phi,v)$.

and

\begin{equation}
\delta B^{\prime}=\frac{\sqrt{\sqrt{1-\frac{V^2}{v^2}}}}{\sqrt{\sqrt{1-\frac{v^2}{c^2}}}}\delta B,
\end{equation}
where $\delta B=\delta B(\phi)$ and $\delta B^{\prime}=\delta B(\phi,v)$. The background fields (shifts)
$\delta\vec E^{\prime}$ and $\delta\vec B^{\prime}$ could be interpreted as being the effective responses to the motion of the particle
that experiments the vacuum-$S_V$, having a dynamical origin.

According to (99) and (100) we find the correction on $\rho^{(2)}_{electromag}=
\frac{1}{2}\epsilon_0(\delta E)^2+\frac{1}{2\mu_0}(\delta B)^2$ (see (33)), namely $\rho^{(2)\prime}=\Psi\rho^{(2)}$. 
 For $v\rightarrow c$, $\delta E^{\prime}$ and $\delta B^{\prime}\rightarrow\infty$, leading to $\rho^{(2)\prime}\rightarrow\infty$ around
the particle, which has to do with the increasing of the inertial mass (energy of the particle). This subject has been well explored
in another paper\cite{41}. On the other hand, for $v\rightarrow V$, $\delta E^{\prime}$ and $\delta B^{\prime}\rightarrow 0$, leading to $\rho^{(2)\prime}\rightarrow 0$, which is responsible for the rapid decreasing of the energy of the particle close to $S_V$.

 It is interesting to notice that, if $V\rightarrow 0$, this implies $\xi=V/c\rightarrow 0$, which leads to $\delta E(\phi,v)=
\delta B(\phi,v)=0$ in eqs.(97) and (98) as indeed expected in absence of the ultra-referential $S_V$ (SR theory).

 New transformations on the fields $\vec E$ and $\vec B$ ($F^{\mu\nu}$) in the space-time of SSR firstly require the preparation of a
4x4 matrix of transformation that recovers Lorentz matrix in the limit $V\rightarrow 0$. A simplified 2x2 matrix for $(1+1)D$ was
obtained in a previous work\cite{12}. Such new transformations plus the transformations (99) and (100) and their implications on the
behavior of the terms $\rho^{(0)}$ and $\rho^{(1)}$ of the eq.(33) will be investigated elsewhere.

\section{\label{sec:level1} Conclusions and prospects}

   We have essentially concluded that the space-time structure where gravity is coupled to the electromagnetic fields, by means of a
 background field for a preferred frame connected to a minimum speed $V$, naturally contains the fundamental ingredients for comprehension
 of the uncertainty principle.

  The present theory has various implications which shall be investigated in coming articles. A new group that is more general than Lorentz
 group will be investigated. We will look for transformations in SSR for the fields $F^{\mu\nu}$ changing
 their forms from a certain non-Galilean reference frame to another one. So we plan to construct a new relativistic electrodynamics with
 the presence of a background field of the ultra-referential $S_V$ and make important applications of it. In a previous work\cite{12}, we
 have already shown that the existence of $S_V$ does not violate the covariance of the Maxwell equations, however we intend to go
 deeper into such subject in coming papers.

  Another deep investigation will propose the development of a new more general relativistic dynamics where the energy of vacuum
 (ultra-referential $S_V$) performs a crucial role for understanding the problem of inertia (the problem of mass anisotropy
 \cite{41}). 

  The quantum non-locality aspect of particles close to $V$ will be also deeply explored.

 The sui-generis nature of the vacuum energy of the ultra-referential $S_V$ has been investigated in another article\cite{12},
 where we have studied its implications in cosmology. We have established a connection between the cosmological constant ($\Lambda$) as a
 cosmological scalar field and the cosmological antigravity starting from the vacuum energy of the
 ultra-referential $S_V$, leading to a new energy-momentum tensor ($T^{\mu\nu}$) for the matter in the presence of such
 sui generis vacuum energy. Hence we have obtained the tiny value of the current cosmological constant
 ($\Lambda\sim 10^{-35}s^{-2}$), which is still not well-understood by quantum field theories because such theories foresee a very high
 value of $\Lambda$, whereas the exact supersymmetric theories foresee a null value for $\Lambda$.

Another relevant investigation is with respect to the problem of the absolute zero temperature in the thermodynamics of a gas. We intend to
 make a connection between the 3rd.law of Thermodynamics and the new dynamics\cite{41} through a relationship between the absolute zero
 temperature and a minimum average speed ($\left<v\right>_N=V$) for $N$ particles of a gas. Since $T=0K$ is thermodynamically
 unattainable, this is due to the impossibility of reaching $\left<v\right>_N=V$ from the dynamics standpoint. This still leads to other
 interesting implications, such as for instance, the Einstein-Bose condensate and the problem of the high refraction index of ultra-cold
 gases, where we intend to estimate the speed of light approaching to $V$ inside the condensate.

 In short, we hope to open up a new research field for various areas of Physics, including condensed matter, quantum field theories,
 cosmology and specially a new exploration for quantum gravity at very low energies.

\end{document}